\documentclass[12pt]{article}
\pdfoutput=1

\usepackage[utf8]{inputenc}
\usepackage[english]{babel}
\usepackage{amsmath,amssymb,amsthm,latexsym}
\usepackage{graphicx}
\usepackage{physics}
\usepackage{float}
\usepackage{booktabs}
\usepackage{xcolor}
\usepackage{cite}
\usepackage{caption}
\usepackage[
   bookmarks=true,
   linktocpage=true,
   colorlinks=true,
   allbordercolors=white, allcolors=blue
]{hyperref}
\usepackage{mciteplus}

\newcommand{\email}[1]{\href{mailto:#1}{#1}}
\newcommand{\MPl}{M_{\rm Pl}}

\graphicspath{{figures/}}

\begin{document}

\title{\LARGE {\bf Predictive hadronic axion models from asymptotically safe quantum gravity}}

\author{\\ Kristjan M\"u\"ursepp$^{a,b}$\footnote{\email{kristjan.muursepp@kbfi.ee}}\,,
Daniele Rizzo$^{a}$\footnote{\email{daniele.rizzo@kbfi.ee}}\\[2ex]
\small {\em $^a$National Institute of Chemical Physics and Biophysics,
R\"avala 10, 10143 Tallinn, Estonia}\\
\small {\em $^b$INFN, Laboratori Nazionali di Frascati, C.P.~13, 00044 Frascati, Italy}\\
}
\date{}
\maketitle

\begin{abstract}
We investigate whether asymptotically safe quantum gravity can render hadronic axion models predictive, considering the standard Kim--Shifman--Vainshtein--Zakharov (KSVZ) model and its extension with an additional $SU(2)_a$ gauge symmetry, which for historical reasons we call the totally asymptotically free (TAF) model. Adopting a heuristic parametrisation of quantum gravity corrections to the running of the matter coupling constants, we determine the ultraviolet fixed points and critical exponents, and identify renormalisation-group trajectories compatible with the measured Standard Model parameters. Requiring all scalar couplings to be associated with irrelevant directions at the ultraviolet fixed point strongly constrains the hypercharge $q$ of the new coloured fermions, excluding the conventional KSVZ benchmark with $E/N=0$. The remaining fixed-point conditions constrain the couplings and heavy-particle spectrum in terms of the hypercharge $q$ and the axion decay constant $f_a$, leading to characteristic bands in the axion mass--photon coupling plane. For the TAF model, we also investigate previously neglected nonperturbative instanton effects from the additional $SU(2)_a$ gauge symmetry and the cosmological implications of stable heavy relics. We show that the exotic coloured fermions are necessarily stable 
in the minimal TAF model, severely constraining the post-inflationary scenario of Peccei--Quinn (PQ) symmetry breaking, and potentially making it cosmologically unviable. Focusing on the pre-inflationary alternative, we derive constraints on reheating from the thermal regeneration of heavy relics and the requirement of avoiding PQ symmetry restoration. Our results connect ultraviolet fixed-point conditions to predictive mass spectra, axion phenomenology and cosmological viability of hadronic axion models.
\end{abstract}
\newpage

\tableofcontents

\setcounter{footnote}{0}

\newpage

\section{Introduction}\label{sec:intro}

The absence of observable charge-parity (CP) violation in the strong interactions remains one of the outstanding puzzles of the Standard Model (SM). In particular, the quantum chromodynamics (QCD) Lagrangian admits a CP-violating topological term characterised by the effective vacuum angle $\bar\theta=\theta+\arg\det M_q$, where $\theta$ is the coefficient of $\frac{g_3^2}{32\pi^2}G^a_{\mu\nu}\tilde G^{a\mu\nu}$ in the Lagrangian and $M_q$ is the quark mass matrix. While its magnitude is constrained to be extremely small, $|\bar\theta|\lesssim 10^{-10}$, by measurements of the neutron electric dipole moment~\cite{Abel:2020pzs}, the SM offers no explanation to this suppression. An elegant solution to the strong CP problem was proposed by Peccei and Quinn (PQ), who introduced a spontaneously broken global $U(1)_{\rm PQ}$ symmetry that dynamically relaxes the effective vacuum angle to its CP-conserving value~\cite{Peccei:1977ur}. The associated pseudo-Nambu--Goldstone boson, the QCD axion, also provides a viable dark matter (DM) candidate~\cite{Preskill:1982cy,Abbott:1982af,Dine:1982ah}. Its mass and low-energy couplings are largely determined by the axion decay constant $f_a$ and the anomaly structure of the underlying theory~\cite{GrillidiCortona:2015jxo}.

Despite the constraints on its low-energy effective interactions, the microscopic realisation of the QCD axion remains largely undetermined. Two familiar possibilities are the Kim--Shifman--Vainshtein--Zakharov (KSVZ) and Dine--Fischler--Srednicki--Zhitnitsky (DFSZ) constructions~\cite{Kim:1979if, Shifman:1979if, Zhitnitsky:1980tq,Dine:1981rt}. In KSVZ-type models, the anomalous axion coupling to gluons is generated by additional heavy coloured fermions that are vector-like under the SM gauge group but chiral under the PQ symmetry~\cite{Kim:1979if,Shifman:1979if}. Their gauge representations, hypercharges and multiplicities are not fixed by the PQ mechanism itself, leaving considerable freedom in the ratio of electromagnetic to colour anomaly coefficients and hence in the axion--photon coupling. Phenomenological and theoretical criteria, including constraints on stable exotic coloured states and perturbativity up to high energies, have therefore been used to identify preferred classes of hadronic axion models~\cite{DiLuzio:2016sbl,DiLuzio:2017pfr,DiLuzio:2020wdo, DiLuzio:2024xnt, Cheek:2023fht}.

The interactions of the scalar field responsible for PQ symmetry breaking are similarly unconstrained by low-energy axion phenomenology. Nevertheless, they can have important implications for Higgs vacuum stability and inflation~\cite{Elias-Miro:2012eoi,Ballesteros:2016euj,Ballesteros:2016xej}, as well as for the cosmological history of the Universe, including PQ symmetry restoration and the evolution of the radial scalar mode~\cite{Moroi:2014mqa,Ballesteros:2021bee,Ringwald:2020vei}. The scalar potential may also give rise to a strongly first-order PQ phase transition, potentially generating observable gravitational-wave signals~\cite{DelleRose:2019pgi,VonHarling:2019rgb,Ghoshal:2020vud,Chiang:2020aui,Yang:2024npd,Benincasa:2026pmw}. These considerations motivate studies of ultraviolet (UV) frameworks capable of constraining the otherwise largely undetermined parameter space of hadronic axion models.

One possible framework is provided by asymptotically safe quantum gravity. Originally proposed by Weinberg, the asymptotic-safety scenario postulates that the renormalisation-group (RG) flow of gravity approaches a UV fixed point with a finite number of relevant directions~\cite{Weinberg:1980gg,Reuter:1996cp}. Such a fixed point would provide a nonperturbative UV completion of gravity, with predictivity arising from the finite dimensionality of the UV critical surface: couplings associated with irrelevant directions are fixed by the requirement that the RG trajectory approaches the fixed point, so their low-energy values can be predicted once the relevant parameters are specified. Functional RG (FRG) studies have provided substantial evidence for such a fixed point, including in gravity--matter systems~\cite{Eichhorn:2018yfc,Pawlowski:2020qer,Eichhorn:2022gku}. For particle physics, the key implication is that quantum-gravitational fluctuations also contribute to the running of the gauge, Yukawa and scalar couplings of the matter sector, and can drive them to either free or interacting fixed points~\cite{Eichhorn:2017ylw,Eichhorn:2017lry,Eichhorn:2017als}. The prototypical example is the Higgs-mass prediction of Shaposhnikov and Wetterich: if the SM coupled to gravity is asymptotically safe, the Higgs quartic coupling is driven to a fixed point at a vanishing value above the Planck scale, which fixes the Higgs mass~\cite{Shaposhnikov:2009pv}. Quantitative predictions, however, remain sensitive to the gravitational truncation, the renormalisation scheme and the treatment of the RG evolution~\cite{Kotlarski:2023mmr}, which motivates the heuristic approach described below.

In the heuristic approach, the gravitational fixed point is encoded above the Planck scale in effective coefficients $f_g$, $f_y$ and $f_\lambda$, which shift the $\beta$-functions of the gauge, Yukawa and quartic couplings linearly in the coupling itself, $\beta_X=\beta_X^{\rm matter}-f_X X$. The coefficients are taken to be universal for each type of interaction, and the gravitational contributions are switched off abruptly below the Planck scale. Rather than computing the coefficients in a specific truncation, one treats them as free parameters constrained by the measured SM couplings, using the FRG results~\cite{Eichhorn:2017ylw,Eichhorn:2017lry,Eichhorn:2017als,Giacometti:2026zrs} as guidance on their sign and size. The stability of the resulting predictions against a smoother decoupling has been tested explicitly in Ref.~\cite{Kotlarski:2023mmr}. This effective approach has been applied to a wide range of SM extensions~\cite{Eichhorn:2017ylw,Eichhorn:2018whv,Reichert:2019car,Alkofer:2020vtb,Kowalska:2020gie,Kowalska:2020zve,Kowalska:2022ypk,Chikkaballi:2022urc,Kotlarski:2023mmr,Chikkaballi:2023cce,Kowalska:2024lnm,Sessolo:2024uqr}, and we also apply the same methodology in this work.

The connection between asymptotic safety and axion physics has been investigated in several earlier studies. First, it was shown that the SM matter content with different new physics extensions, including those similar to KSVZ models, can be compatible with asymptotically safe gravitational fixed points~\cite{Dona:2013qba}. More directly, the Higgs-mass prediction in a KSVZ-inspired model was studied in Ref.~\cite{Kwapisz:2019wrl}, where it was concluded that additional coloured fermions induce only a small correction when their effect on the SM running is mediated primarily through the QCD coupling. Related studies have examined the impact of Higgs-portal interactions on the Higgs-mass prediction in asymptotically safe dark-sector extensions~\cite{Eichhorn:2021tsx}. A complementary approach was pursued in Ref.~\cite{deBrito:2021akp}, which considered fundamental axion-like particles coupled to asymptotically safe gravity. It was found that gravitational fluctuations can substantially modify the UV scaling of the dimension-five axion--photon interaction, although the gravitational strength required for a UV completion is in tension with the weak-gravity bound. These investigations leave open the question of how gravitational fixed-point conditions constrain the interactions and phenomenology of a renormalisable PQ-breaking sector.

Other approaches have sought UV completeness within the axion sector itself. The totally asymptotically free (TAF) QCD axion model was constructed in Ref.~\cite{Salvio:2020prd}, with all the matter couplings, aside from those of the SM, vanishing at arbitrarily high energies. The model introduces an additional $SU(2)_a$ gauge symmetry, a complex scalar in its adjoint representation and new coloured fermions, with spontaneous symmetry breaking leaving an unbroken Abelian gauge symmetry associated with a massless dark photon. The first order phase transitions and the associated gravitational wave signals that result from post-inflationary PQ breaking in this model were subsequently studied in~\cite{Ghoshal:2020vud}. A different realisation was proposed in~\cite{Ishida:2021avk}, which considered a composite axion from a walking gauge theory exhibiting an interacting UV fixed point. These constructions illustrate alternative routes towards UV-complete axion models without relying on asymptotically safe quantum gravity.

In this work, we investigate the predictive implications of asymptotically safe quantum gravity for two renormalisable hadronic axion constructions: the conventional KSVZ model and the TAF model of Ref.~\cite{Salvio:2020prd}. We retain the name of the latter, although in the presence of gravitational corrections some of its couplings approach interacting rather than Gaussian UV fixed points. Once the coefficients $f_g$, $f_y$ and $f_\lambda$ are fixed by the measured SM parameters, the UV fixed-point conditions restrict the interactions of the new PQ sector. We determine the resulting fixed points and critical exponents analytically at one loop and investigate the corresponding RG trajectories, including higher-order running and threshold matching.

A central result of our analysis is that requiring the Higgs quartic coupling to be fixed by an irrelevant UV direction, while reproducing its measured low-energy value, strongly constrains the hypercharge $q$ of the new coloured fermions. Unlike the earlier KSVZ analysis of Ref.~\cite{Kwapisz:2019wrl}, we systematically investigate the effect of non-zero fermion hypercharge on the electroweak gauge running. We find that the viable fixed-point branches require non-vanishing hypercharge, excluding the conventional hypercharge-neutral KSVZ benchmark within our assumptions. Since the electromagnetic-to-colour anomaly ratio is $E/N=6q^2$ in both models, these restrictions translate into characteristic regions in the axion mass--photon coupling plane. Furthermore, the fixed-point conditions constrain the Yukawa and scalar interactions, relating the spectrum of heavy particles to $q$ and $f_a$. For the TAF model, we also investigate how gravitational corrections modify its fixed-point structure, with the additional gauge interaction driving radiative PQ symmetry breaking through the Coleman--Weinberg (CW) mechanism~\cite{Coleman:1973jx}. We also show that the Higgs--PQ portal coupling is irrelevant and vanishes at the one-loop fixed point, thereby avoiding large threshold corrections to the Higgs mass parameter at this order. At two loops, the portal coupling is regenerated, and we estimate the resulting fine-tuning. We further find that, although a deeper minimum of the Higgs effective potential may develop at large field values due to the running of $\lambda_H$ towards negative values, the tunnelling rate from the electroweak vacuum is further suppressed relative to the SM, resulting in an even longer vacuum lifetime, well beyond the age of the Universe.

Beyond the RG analysis, we examine the cosmological consistency of the resulting models, with particular attention to the stability and abundance of their new heavy states. In the minimal $SU(2)_a$ construction, the exotic coloured fermions are stable, leading to severe relic-abundance constraints that effectively exclude the standard post-inflationary PQ-breaking scenario in the minimal model. We therefore investigate the pre-inflationary alternative, in which an initial population of heavy relics is diluted by inflation but can be regenerated during reheating. We estimate the resulting constraints from thermal production, and the requirement of symmetry non-restoration after reheating, deriving an upper bound on the reheating temperature
as a function of the axion decay constant and the
hypercharge of the exotic fermions. We also examine the phenomenology of the additional $SU(2)_a$ gauge sector, including its massive vector bosons, the surviving dark photon and nonperturbative instanton effects on the axion potential, which were not studied in the original formulation of the model. Throughout, we assume the existence of a sufficiently accurate PQ symmetry, whose protection against higher-dimensional symmetry-violating operators is not established by the asymptotic safety of the renormalisable matter interactions alone.

The paper is organised as follows. In Sec.~\ref{sec:models} we introduce the two hadronic axion models and their particle content. In Sec.~\ref{sec:asqg} we describe the asymptotic-safety framework and determine the UV fixed points and critical exponents. In Sec.~\ref{sec:results} we identify predictive RG trajectories compatible with the measured SM parameters and investigate their implications for the heavy-particle spectrum, axion phenomenology and cosmological viability. We conclude with a summary of our findings.

%%%%%%%%%%%%%%%%%%%%%%%%%%%%%%%%%%%%%%%%%%%%%%%%%%%%%%%%%%%%%%%%%%%%%%%%%%%%%%%%%%%%%%%%%%%%%%%%%%%%%%%%%%%%%
\section{QCD axion models\label{sec:models}}
%%%%%%%%%%%%%%%%%%%%%%%%%%%%%%%%%%%%%%%%%%%%%%%%%%%%%%%%%%%%%%%%%%%%%%%%%%%%%%%%%%%%%%%%%%%%%%%%%%%%%%%%%%%%%

We consider two hadronic axion models, in which the anomalous axion-gluon coupling is generated by the presence of coloured fermions. In both models, a complex scalar that is a singlet under
the SM gauge group acquires a vacuum expectation value
(vev), giving rise to a pseudo-Nambu--Goldstone boson $a$,
the axion, whose anomalous coupling $a\,G\widetilde{G}$ to
the gluon field strength tensor provides a dynamical
solution to the strong CP problem. The main difference between the two models is the presence of an
additional $SU(2)_a$ gauge symmetry in the second model, which enables us to consider radiative PQ symmetry breaking instead of tree-level
symmetry breaking.

In both models
the new quarks are $SU(3)_{c}$ triplets with non-zero hypercharge $q$. In contrast, the PQ scalar transforms as a singlet under $SU(3)_c$ and has zero hypercharge under $U(1)_{Y}$. The quantum numbers of both the new fields and the relevant SM fields are compactly summarised in Table~\ref{tab:fields}.

\begin{table}[t]\centering
\begin{tabular}{llccccc}
\toprule
& Field & $SU(3)_c$ & $SU(2)_L$ & $U(1)_Y$ & $SU(2)_a$ & $U(1)_{\rm PQ}$ \\
\midrule
SM   & $Q_L$  & $\mathbf{3}$ & $\mathbf{2}$ & $1/6$ & $\mathbf{1}$ & $0$ \\
     & $t_R$  & $\mathbf{3}$ & $\mathbf{1}$ & $2/3$ & $\mathbf{1}$ & $0$ \\
     & $H$    & $\mathbf{1}$ & $\mathbf{2}$ & $1/2$ & $\mathbf{1}$ & $0$ \\
\midrule
KSVZ & $\phi$      & $\mathbf{1}$ & $\mathbf{1}$ & $0$ & $\mathbf{1}$ & $1$ \\
     & $\psi_L$      & $\mathbf{3}$ & $\mathbf{1}$ & $q$ & $\mathbf{1}$ & $0$ \\
     & $\psi_R$  & $\mathbf{3}$ & $\mathbf{1}$ & $q$ & $\mathbf{1}$ & $-1$ \\
\midrule
TAF  & $A$         & $\mathbf{1}$ & $\mathbf{1}$ & $0$ & $\mathbf{3}$ & $1$ \\
     & $\psi_L$      & $\mathbf{3}$ & $\mathbf{1}$ & $q$ & $\mathbf{2}$ & $0$ \\
     & $\psi_R$  & $\mathbf{3}$ & $\mathbf{1}$ & $q$ & $\mathbf{2}$ & $-1$ \\
\bottomrule
\end{tabular}
\caption{Field content and charge assignments of the two hadronic
axion models considered in this work, together with the relevant
SM fields.}\label{tab:fields}
\end{table}

\subsection{The KSVZ axion\label{sec:model_ksvz}}

We begin by studying the well-known
KSVZ model. To that end, we augment the SM with a complex scalar $\phi$, singlet under the SM gauge group, and one vector-like colour
triplet $\psi$. The Yukawa sector and the scalar potential are given by
\begin{align}
  -\mathcal{L}_{\rm Yuk}&=y_t\,\bar Q_L\widetilde H\,t_R+\,\sqrt2\,y_\psi\,\phi\,\bar\psi_L\psi_R+{\rm h.c.},\label{eq:ksvz_yuk}\\
  V&=-\mu_H^2\,H^\dagger H+\lambda_H(H^\dagger H)^2-\mu_\phi^2\,|\phi|^2+\lambda_\phi|\phi|^4
     +\lambda_{H\phi}\,H^\dagger H\,|\phi|^2,\label{eq:ksvz_V}
\end{align}
where $\widetilde H=i\sigma_2H^*$ and we retain only the dominant Yukawa coupling $y_t$ for simplicity.
The vev $\langle\phi\rangle=v_\phi/\sqrt2$ breaks the PQ symmetry. Writing
$\phi=(v_\phi+\rho)\,e^{ia/v_\phi}/\sqrt2$, the heavy states are the quark $\psi$ and the radial mode $\rho$, with
\begin{equation}\label{eq:ksvz_masses}
  m_\psi=y_\psi\,v_\phi,\qquad m_\rho^2=2\lambda_\phi\,v_\phi^2 ,
\end{equation}
where the radial mode mass receives relative corrections of order $v_{\rm EW}^2/v_\phi^2$
from its mixing with the Higgs, with $v_{\rm EW} = 246\, {\rm GeV}$ denoting the Higgs vev. The Yukawa term involving $\phi$ then becomes
\begin{equation}
     m_\psi \left(1 + \frac{\rho}{v_{\phi}}\right) e^{i a/v_{\phi}} \bar\psi_L \psi_R + \text{h.c.}
\end{equation}
The phase appearing in the mass term can then be removed by a chiral rotation 
\begin{equation}
    \psi_L \rightarrow \psi_L, \quad \psi_R \rightarrow e^{-i a/(v_\phi)} \psi_R,
    \label{eq:KSVZaxiontransf}
\end{equation}
whose anomalous Jacobian generates the axion--gluon interaction
\begin{equation}
    \Delta\mathcal L =
    \frac{a}{v_\phi}\mathcal A_{SU(3)_c}
    \frac{g_3^2}{16\pi^2}
    G^A_{\mu\nu}\widetilde G^{A\mu\nu},
\end{equation}
with $A$ denoting the index of the adjoint representation of $SU(3)_c$ and the anomaly coefficient given by
\begin{equation}
\mathcal A_{SU(3)_c}
=(X_L - X_R)
T_{SU(3)_c}(R_\psi)
\dim_{\psi}^G,
\end{equation}
where $X_{L}, X_R$ denote the PQ charges of $\psi_L, \psi_R$ respectively. The Dynkin index $T_{SU(3)_c}$ is given in terms of the trace constructed from the corresponding generators
\begin{equation}
    \Tr{ T^A T^B} = T_{SU(3)_c}(R_\psi) \delta^{AB}
\end{equation}
and $\dim_\psi^G$ denotes the multiplicity of $\psi$ under other gauge groups. In the KSVZ model we have $T_{SU(3)_c}(\mathbf{3}) = \frac{1}{2}$ and $\dim_\psi^G = 1$, and so $\mathcal{A}_{SU(3)_c} = \frac{1}{2}$. 

The anomalous coupling of the axion to the gluons generates a nonperturbative
potential for the axion~\cite{GrillidiCortona:2015jxo,Borsanyi:2016ksw}. In the cosine
approximation it can be written as
\begin{equation}
    V_{\rm QCD}(a)
    =
    -\chi_{\rm QCD}
    \cos\left(
        N\frac{a}{v_{\phi}}
        +\bar\theta
    \right),
\end{equation}
where $N=2\mathcal A_{SU(3)_c}$ is the colour anomaly
coefficient, which in the KSVZ model coincides with the
physical domain-wall number, $N_{\rm DW}=N=1$, and $\chi_{\rm QCD}^{1/4}\simeq 75.5~{\rm MeV}$ is the zero-temperature QCD topological susceptibility~\cite{GrillidiCortona:2015jxo}. The axion decay constant and mass are thus given by
\begin{equation}
    f_a\equiv\frac{v_\phi}{N},\qquad
    m_a^2=\frac{\chi_{\rm QCD}}{f_a^2}.
\end{equation}

The transformation of the fermion fields in Eq.~\eqref{eq:KSVZaxiontransf} also results in the anomalous axion-photon coupling through the $U(1)_{\rm PQ} \times U(1)_Y^2$ anomaly. The low energy coupling is given by~\cite{GrillidiCortona:2015jxo},
\begin{equation}
    g_{a \gamma} = \frac{\alpha}{2 \pi f_a } \left[ \frac{E}{N} - 1.92 \right],
\end{equation}
where $E$ is the anomaly coefficient of the $U(1)_{\rm PQ}\times U(1)_Y^2$ anomaly, defined as
\begin{equation}
    E\equiv 2 \mathcal{A}_{U(1)_Y} = 6 (X_L - X_R) q^2 = 6 q^2.   
\end{equation}

Previous works have shown that specific hypercharge assignments can be phenomenologically preferred, either by the requirement that the heavy fermions decay in the post-inflationary scenario or by the absence of Landau poles below the Planck scale~\cite{DiLuzio:2016sbl,DiLuzio:2017pfr}. In this work, we instead consider the pre-inflationary scenario, in which primordial heavy-fermion abundances are diluted
by inflation and their subsequent regeneration during
reheating must be sufficiently suppressed to avoid overproduction. Consequently, we vary the hypercharge $q$ as a free parameter in our numerical studies. The post-inflationary KSVZ scenario, where the hypercharge is restricted, is the subject of a follow-up work currently in preparation~\cite{MuurseppRizzo:PostInflationary}.

Since $\psi$ is an $SU(2)_L$ singlet, its electric charge equals its hypercharge $q$, whereas every colour-triplet state made of SM fields has charge $-1/3$ modulo integers. For $q\notin-1/3+\mathbb Z$, hadrons containing a single
$\psi$ are therefore fractionally charged and cannot
decay exclusively into SM particles. Instead, for $q\in-1/3+\mathbb Z$, $\psi$ can decay into SM states through renormalisable mixing with $d_R$ or $u_R$ for $q=-1/3$ and $2/3$, or through dimension-six operators for $q=5/3$ and $8/3$. We do not include these additional couplings in our fixed-point analysis. They would receive their own gravitational corrections, which effectively modify the gravitational coefficients. Fixed-point predictions have been shown to be robust against comparable modifications of the heuristic setup, with uncertainties below one per cent for the gauge couplings and of at most a few to ten per cent for Yukawa couplings of natural size~\cite{Kotlarski:2023mmr}. We therefore do not expect these couplings to affect our results. In addition, the renormalisable mixings concern only $q=-1/3$ and $2/3$, which lie below the range allowed by predictivity, $q\gtrsim1.3$, as we shall see below.

\subsection{The TAF axion\label{sec:model_taf}}

The second model we consider is the TAF axion model of
Ref.~\cite{Salvio:2020prd}, which extends the KSVZ axion
sector by an additional $SU(2)_a$ gauge symmetry and allows
for radiative PQ symmetry breaking through dimensional
transmutation.

The PQ scalar $A$ is a complex field transforming in the adjoint representation of $SU(2)_a$,
\begin{equation}
    A=A^a\frac{\sigma^a}{2},\qquad a=1,2,3,
\end{equation}
and therefore contains three complex, or equivalently six real, scalar degrees of freedom.
The kinetic term of $A$ is given by
\begin{equation}
    \mathcal L_{\rm kin}
    =\Tr{(D_\mu A)^\dagger D^\mu A},
    \qquad
    D_\mu A=\partial_\mu A-ig_a[V_\mu,A],
\end{equation}
where $V_\mu=V_\mu^a\sigma^a/2$ and $g_a$ is the
$SU(2)_a$ gauge coupling.

The vector-like coloured fermion $\psi$ transforms instead in the fundamental representation of $SU(2)_a$. Its transformation properties under $SU(3)_c\times U(1)_Y$ mimic those of the heavy fermion in the KSVZ model, while all SM fields are singlets under $SU(2)_a$. In particular, the two components of the $SU(2)_a$ multiplet carry hypercharge $q$. The relevant Yukawa interactions and scalar potential are
\begin{align}
  -\mathcal{L}_{\rm Yuk}
  &=
  y_t\,\bar Q_L\widetilde H\,t_R
  +y_\psi\,\bar\psi_L A\psi_R
  +{\rm h.c.},
  \label{eq:taf_yuk}\\
  V
  &=
  -\mu_H^2\,H^\dagger H
  +\lambda_H(H^\dagger H)^2
  -\mu_A^2\,{\rm Tr}(A^\dagger A)
  +\lambda_1\,{\rm Tr}^2(A^\dagger A)
  \nonumber\\
  &\quad
  +\lambda_2\,|{\rm Tr}(AA)|^2
  +\lambda_{HA}\,H^\dagger H\,{\rm Tr}(A^\dagger A).
  \label{eq:taf_V}
\end{align}

We consider the regime in which the explicit mass
parameters in the scalar potential are negligible
for the dynamics of PQ symmetry breaking.
In particular, we assume
$|\mu_A^2|\ll\bar\beta v_A^2$,
where $\bar\beta$ and $v_A$ are defined below,
such that radiative corrections dominate the
generation of $f_a$. The Higgs mass
parameter is retained to account for electroweak
symmetry breaking at lower energies.

As we will discuss in the next section, for predictive axion models, the vanishing Higgs--PQ scalar portal coupling $\lambda_{HA}=0$
corresponds to an irrelevant fixed point. At one-loop level, the SM and PQ scalar sectors can therefore be kept decoupled throughout the RG evolution without spoiling asymptotic safety. At two loops, however, a non-zero $\lambda_{HA}$ is generated radiatively (see Eq.~\eqref{eq:portal2L}), so in the numerical analysis we keep it on the corresponding trajectory instead of setting it to zero, see Sec.~\ref{sec:asqg_portal}. Since the PQ transition takes place far above the electroweak scale, it can then be studied along the trajectory
\begin{equation}
    H=0.
\end{equation}
The radiative generation of $f_a$ consequently reduces, at this order, to the application of the CW mechanism~\cite{Coleman:1973jx,Gildener:1976ih} involving the field $A$ alone. In contrast, the Higgs vev is generated in the usual way without the need for the CW mechanism.

At the renormalisation scale $\mu_{\rm PQ}$, defined by
\begin{equation}
    \lambda_1(\mu_{\rm PQ})+\lambda_2(\mu_{\rm PQ})=0,
    \qquad \lambda_1(\mu_{\rm PQ})>0,
    \label{eq:taf_flat}
\end{equation}
the tree-level potential develops a flat direction.
Up to $SU(2)_a$ gauge and PQ transformations, the scalar
background along this direction can be parametrised as
\begin{equation}
    A=\frac{\varphi}{2}\sigma^1,
    \label{eq:taf_flatdir}
\end{equation}
along which the tree-level potential reduces to
\begin{equation}
    V_0(\varphi)=\frac{\lambda_1+\lambda_2}{4}\varphi^4.
    \label{eq:taf_flatV}
\end{equation}
Equation~\eqref{eq:taf_flat} therefore does not by itself generate a non-zero vev. Rather, it implies that the classical potential is flat along the $\varphi$ direction at the scale $\mu_{\rm PQ}$.

The flat direction is lifted by radiative corrections.
In the leading-logarithmic approximation, the one-loop
effective potential reads~\cite{Ghoshal:2020vud}
\begin{equation}
  V_{\rm CW}(\varphi)
  =
  \frac{\bar\beta}{4}
  \left(
      \ln\frac{\varphi}{v_A}
      -\frac14
  \right)\varphi^4,
  \label{eq:taf_VCW}
\end{equation}
where
\begin{equation}
  \bar\beta
  =
  \frac{
      12g_a^4
      +16\lambda_1^2
      -3y_\psi^4
  }{16\pi^2}
  \bigg|_{\mu_{\rm PQ}}.
  \label{eq:taf_beta}
\end{equation}
For $\bar\beta>0$, the quantum-corrected potential develops a minimum at a non-zero field value,
\begin{equation}
    \langle\varphi\rangle=v_A,
\end{equation}
with
\begin{equation}
    v_A= e^{-1/4}\mu_{\rm PQ}.
    \label{eq:taf_vmu}
\end{equation}

The heavy-particle spectrum is governed by the simultaneous breaking of the global $U(1)_{\rm PQ}$ and $SU(2)_a$ symmetries, and is given by
\begin{align}
    m_{V}&=g_av_A,
    &m_\psi&=\frac{y_\psi v_A}{2},\\
    m_{S_1}^2&=\bar\beta v_A^2,
    &m_{S^\pm}^2&
=(\lambda_1-\lambda_2)v_A^2,
\end{align}
where the two massive gauge bosons $V^\pm$ both have mass $m_V$,
$S_1$ is the pseudo-dilaton associated
with the spontaneous breaking of approximate
scale invariance, while the two remaining scalar modes $S_2$ and $S_3$ combine into a complex scalar
with charged components $S^\pm$ under the residual
$U(1)_a$ gauge symmetry.
Of the three Goldstone modes arising from $A$,
two are eaten by $V^\pm$, while the third
corresponds to the axion.
The unbroken $U(1)_a$ gauge symmetry gives
rise to a massless dark photon
\cite{Salvio:2020prd}.

Along the vacuum manifold, the scalar field $A$ can be
parametrised as
\begin{equation}
    A=\frac{v_A}{2}e^{ia/v_A}\sigma^1,
\end{equation}
where $a$ is canonically normalised.
The evaluation of the $U(1)_{\rm PQ}\times SU(3)_c^2$ anomaly coefficient proceeds in close analogy to the calculation in Sec.~\ref{sec:model_ksvz}, with the only difference being that here $\dim_{\psi}^{SU(2)_a} = 2$. Thus we obtain 
\begin{equation}
    N= 2\mathcal{A}_{SU(3)_c} = 2,
    \qquad
    f_a=\frac{v_A}{N}=\frac{e^{-1/4}}{2}\,\mu_{\rm PQ},
    \label{eq:TAF_N_and_fa_definition}
\end{equation}
and the electromagnetic-to-colour anomaly ratio is
\begin{equation}\label{eq:taf_EN}
    \frac EN=6q^2 ,
\end{equation}
as in the KSVZ model. Although $N=2$, the physical domain-wall number is
$N_{\rm DW}=1$, since the two minima of the QCD-induced
potential are related by an $SU(2)_a$ gauge transformation
mapping $A$ to $-A$. The domain-wall problem is therefore
avoided through the Lazarides--Shafi
mechanism~\cite{Lazarides:1982tw,Sato:2018nqy}.

\subsubsection{\texorpdfstring{$SU(2)_a$}{SU(2)\_a} instanton contributions}\label{sec:taf_instantons}

The PQ symmetry is anomalous not only under QCD but also
under $SU(2)_a$. The corresponding anomaly coefficient
$N_a$ is given by
\begin{equation}
    N_a = 2\mathcal{A}_{SU(2)_a}
    = 2(X_L-X_R)\,T_{SU(2)_a}(\mathbf{2})\,
    \dim_\psi^{SU(3)_c}=3,
\end{equation}
where $T_{SU(2)_a}(\mathbf{2})=1/2$. As in QCD, the
$SU(2)_a$ gauge sector admits a topological $\theta$ term
with an independent vacuum angle $\theta_a$. The $SU(2)_a$ anomaly would then naively allow a charge-one instanton contribution of the form
\begin{equation}
    V_{Q=1}(a)\propto -\Lambda_{a,1}^4\cos\!\left(\frac{3a}{v_A}+\theta_a\right).
\end{equation}
However, as for the QCD-induced potential, the vacuum configurations $A$ and $-A$ are related by an $SU(2)_a$ gauge transformation, implying the identification
\begin{equation}
    \frac{a}{v_A} \sim \frac{a}{v_A}+\pi
\end{equation}
in the physical vacuum manifold~\cite{Sato:2018nqy}. The effective potential must therefore satisfy
\begin{equation}
    V(a+\pi v_A)=V(a).
\end{equation}
Since
\begin{equation}
    \cos\!\left[3\left(\frac{a}{v_A}+\pi\right)+\theta_a\right]
    =-\cos\!\left(\frac{3a}{v_A}+\theta_a\right),
\end{equation}
the charge-one instanton contribution is forbidden. More generally, the same argument excludes
contributions from all sectors of odd
topological charge.

The same result follows from the fermionic zero modes of the $SU(2)_a$ instanton~\cite{tHooft:1976snw}. The $\psi$ doublet carries three colours, so the 't~Hooft vertex requires three insertions of the Yukawa coupling $\bar\psi_L A\psi_R$, which must combine into an $SU(2)_a$ invariant built from the adjoint field $A$. The only cubic invariant is $\mathrm{Tr}\,A^3$, but it vanishes identically, because the cube of a traceless $2\times2$ matrix is proportional to the matrix itself and therefore has zero trace. 
The one-instanton contribution therefore vanishes, in agreement with the symmetry argument above. 

The leading symmetry-allowed contribution
arises from the topological-charge-two sector and
takes the form
\begin{equation}
    V_a(a)
    =
    -\Lambda_a^4
    \cos\!\left(6 \frac{a}{v_A}+2\theta_a\right).
    \label{eq:TAF_VaSU(2)}
\end{equation}
Following the semiclassical approach of
Ref.~\cite{Csaki:2023ziz}, we adopt the schematic estimate
\begin{equation}
    \Lambda_a^4=
    \mathcal K_2\,y_\psi^6(v_A)\,v_A^4
    \left(\frac{8\pi^2}{g_a^2(v_A)}\right)^8
    \exp\left[-\frac{16\pi^2}{g_a^2(v_A)}\right],
    \label{eq:TAF_Lambda}
\end{equation}
where the exponential reflects the classical action of a
topological-charge-two configuration, while the factor
$(8\pi^2/g_a^2)^8$ arises from the Jacobian associated
with the transformation from gauge-field fluctuations
to the sixteen bosonic collective coordinates of the
charge-two $SU(2)$ instanton sector. The factor $y_\psi^6$
originates from the six Yukawa insertions required to saturate
the twelve fermionic zero modes. Finally, the dimensionless
coefficient $\mathcal K_2$ parametrises the remaining
functional determinants and collective-coordinate
integrals, including correlations between the instantons.

The coefficient $\mathcal K_2$ may incorporate
substantial numerical suppressions arising from
the instanton measure, the integration over
instanton sizes and the overlap of fermionic
zero modes. In the absence of an explicit
calculation of the connected charge-two
contribution, we adopt $\mathcal K_2=10^{-15}$
as an illustrative benchmark. As we demonstrate
below, the resulting upper bound on $g_a(v_A)$
is relatively insensitive to the precise value
of this prefactor.

Including the QCD contribution, the axion potential reads
\begin{equation}
    V(a)\simeq
    -\chi_{\rm QCD}\cos\left(\frac{2a}{v_A}+\bar\theta\right)
    -\Lambda_a^4\cos\left(\frac{6a}{v_A}+2\theta_a\right).
\end{equation}
Minimising the potential in the limit
$\Lambda_a^4\ll\chi_{\rm QCD}$ yields a residual strong CP phase
\begin{equation}
    \theta_{\rm eff}\equiv
    \frac{2\langle a\rangle}{v_A}+\bar\theta
    \simeq-\frac{3\Lambda_a^4}{\chi_{\rm QCD}}
    \sin\left(2\theta_a-3\bar\theta\right).
\end{equation}

Since $\theta_a$ is a CP phase independent of $\bar\theta$, there is no reason for the sine to be small, and the bound $|\theta_{\rm eff}|\lesssim10^{-10}$ requires
\begin{equation}
\frac{3\Lambda_a^4}{\chi_{\rm QCD}}<10^{-10}.
\end{equation}
With Eq.~\eqref{eq:TAF_Lambda}, this becomes
\begin{equation}
2\,\frac{8\pi^2}{g_a^2(v_A)}-8\ln\frac{8\pi^2}{g_a^2(v_A)}>\ln\!\left(\frac{3\mathcal{K}_2\,y_\psi^6(v_A)\,v_A^4}{\chi_{\rm QCD}}\right)+10\ln10 ,
\label{eq:SU(2)InstantonBound}
\end{equation}
which gives an upper bound on $g_a(v_A)$.

The sensitivity of this bound to the precise value of $\mathcal K_2$ can be quantified by differentiating
Eq.~\eqref{eq:SU(2)InstantonBound} at saturation
with respect to $\ln\mathcal K_2$, keeping $v_A$
and $y_\psi$ fixed. We obtain
\begin{equation}
    \frac{\partial\ln g_a^{\rm max}}
    {\partial\ln\mathcal K_2}
    =
    -\frac{(g_a^{\rm max})^2}
    {32\pi^2-16(g_a^{\rm max})^2}
    \simeq
    -\frac{(g_a^{\rm max})^2}{32\pi^2},
\end{equation}
where the last approximation holds for $g_a < \pi$. Consequently, varying
$\mathcal K_2$ by five orders of magnitude
in either direction,
$\Delta\log_{10}\mathcal K_2=\pm5$,
changes the upper bound on $g_a$ by only
approximately $4\%$ for $g_a^{\rm max}\simeq1$.
The constraint on the gauge coupling is
therefore considerably less sensitive to the
unknown numerical prefactor than the instanton
amplitude itself.

\subsubsection{Stability of the exotic fermions}\label{sec:taf_fermion_stability}

Unlike in the KSVZ model, the stability of the lightest exotic coloured state in the minimal TAF model does not depend on its hypercharge. Under the nontrivial element of the centre $\mathbb Z_2\subset SU(2)_a$, the fundamental fermions $\psi_{L,R}$ change sign,
whereas the adjoint scalar, gauge fields and all SM fields remain invariant. Consequently, every
gauge-invariant local operator must contain an even
number of fundamental fermion fields. This forbids
the decay of the lightest centre-charged exotic state
into SM particles, adjoint scalars or gauge bosons,
independently of the operator dimension. QCD confinement
does not remove this protection, implying the existence
of stable exotic hadrons in the minimal model.

One possibility to evade this conclusion would be to introduce additional, lighter $SU(2)_a$ fundamental fermions into which the original exotic quarks could decay. However, the simplest such extension does not resolve the problem. For example, with the PQ charge assignments $X_{\psi_L}=0$, $X_{\psi_R}=-1$ (Table~\ref{tab:fields}) and $X_A=1$, an additional fundamental $\chi_L$ with $X_{\chi_L}=0$ permits the Yukawa interaction
\begin{equation}
\mathcal{L}\supset
-y_{\chi \psi}\overline{\chi}_L A \psi_R+\mathrm{h.c.}
\end{equation}
Since $A$ is a colour singlet, gauge invariance requires $\chi_L$ to carry the same SM gauge quantum numbers as $\psi_R$. An anomaly-free extension can be obtained by introducing a corresponding right-handed partner $\chi_R$. The new Yukawa interaction then generates mass mixing among coloured fermions after PQ breaking, but does not remove the conserved centre charge. Consequently, irrespective of the resulting
fermion mass hierarchy, the lightest exotic quark remains protected against decay into
SM particles, implying the existence of stable
exotic hadrons after QCD confinement.

Additional difficulties arise from the PQ charge assignments. For instance, choosing $X_{\chi_R}=1$ allows a Yukawa coupling $\overline{\chi}_R A\chi_L$, generating masses of order $f_a$ for generic Yukawa couplings. However, in this case the contributions of $\psi$ and $\chi$ to the PQ--QCD anomaly cancel, spoiling the axion solution to the strong CP problem unless further anomalous fermions are introduced. Avoiding stable exotic hadrons would therefore require a more substantial extension, involving lighter colour-singlet states carrying the nontrivial centre charge and additional interactions allowing the coloured fermions to decay into them. Such extensions would also modify the RG evolution and potentially the UV fixed-point structure considered here.

The resulting stable coloured relics severely constrain
the post-inflationary PQ-breaking scenario~\cite{DiLuzio:2017pfr}, motivating our focus on
pre-inflationary PQ breaking, in which primordial relics
are diluted by inflation and their subsequent regeneration
can be suppressed by a sufficiently low reheating
temperature. In this scenario, gravitational waves
generated during the PQ phase transition would also
be diluted by inflation.

%%%%%%%%%%%%%%%%%%%%%%%%%%%%%%%%%%%%%%%%%%%%%%%%%%%%%%%%%%%%%%%%%%%%%%%%%%%%%%%%%%%%%%%%%%%%%%%%%%%%%%%%%%%%
\section{Asymptotically safe gravity and the UV fixed point\label{sec:asqg}}
%%%%%%%%%%%%%%%%%%%%%%%%%%%%%%%%%%%%%%%%%%%%%%%%%%%%%%%%%%%%%%%%%%%%%%%%%%%%%%%%%%%%%%%%%%%%%%%%%%%%%%%%%%%%%

In this section we present the framework of the asymptotic safety analysis for the KSVZ and TAF models. To illustrate the main techniques, we consider only the simplest representation $(\mathbf{3},\mathbf{1},q)$ of $SU(3)_c\times SU(2)_L\times U(1)_Y$ for the heavy quarks. Firstly, in Sec.~\ref{sec:asqg_heuristic} we discuss how quantum gravity modifies the $\beta$-functions of the matter couplings as well as the fixed points and critical exponents that follow from the couplings. Then in Secs.~\ref{sec:asqg_gauge}--\ref{sec:asqg_pq} we compute analytically, at one loop, the UV fixed points and the critical exponents of the gauge, Yukawa, Higgs and PQ-sector couplings. In particular, in Sec.~\ref{sec:asqg_sheet} we derive the condition under which the measured Higgs quartic coupling is reproduced from its irrelevant UV direction, which fixes the gravitational coefficient $f_\lambda$, which is subsequently applied to the discussion of the portal coupling and the hierarchy problem in Sec.~\ref{sec:asqg_portal}.

\subsection{Gravitational contributions to the \texorpdfstring{$\beta$}{beta}-functions and fixed points\label{sec:asqg_heuristic}}

We assume that above the Planck scale $\MPl = (8\pi G_N)^{-1/2}\simeq2.435\times10^{18}\,\mathrm{GeV}$ all the fields present in our models couple to quantum gravity, which develops an
interacting fixed point~\cite{Weinberg:1980gg,Reuter:1996cp,Lauscher:2001ya,Reuter:2001ag,Manrique:2011jc} (for a review, see Ref.~\cite{Eichhorn:2026uqj}) and thereby induces a fixed point in the flow of the dimensionless matter
couplings. In the heuristic approach that we follow, the matter $\beta$-functions then receive parametric corrections in the trans-Planckian regime (see Sec.~VIII~E of Ref.~\cite{Eichhorn:2026uqj} for a review of this parametrisation),
\begin{equation}\label{eq:grav}
  \beta_{g}=\beta_{g}^{\rm matter}-f_g\,g,\qquad
  \beta_{y}=\beta_{y}^{\rm matter}-f_y\,y,\qquad
  \beta_{\lambda}=\beta_{\lambda}^{\rm matter}-f_\lambda\,\lambda\qquad(\mu>\MPl),
\end{equation}
where $\beta_X\equiv dX/d\ln\mu$, $g$, $y$ and $\lambda$ stand for all gauge, Yukawa and quartic couplings, and
$\beta^{\rm matter}$ are the $\beta$-functions of Appendix~\ref{app:betas}. At leading order (LO) in the matter couplings, dressing a matter vertex by a graviton exchange, simply rescales it, and so the gravitational contributions are linear in the couplings. The corresponding coefficients $f_g, f_y, f_\lambda$ are functions of the dimensionless Newton
coupling and cosmological constant at the gravitational fixed point, and are therefore constant in the fixed-point regime.

In this approach, the coefficients $f_g$, $f_y$ and $f_\lambda$ are taken to be
universal: they depend on the type of interaction but not on the quantum numbers of the fields involved, so that $f_g$
is universal for all gauge couplings, $f_y$ for all Yukawa couplings and $f_\lambda$ for all quartic couplings.
This universality follows from the fact that gravity is blind to internal symmetries~\cite{Eichhorn:2026uqj}, and it holds as long as additional contributions, for example from non-minimal couplings, are neglected.
Below the Planck scale, the gravitational contributions decouple, since the dimensionful Newton constant suppresses them by $\mu^2/\MPl^2$. This fast suppression is well approximated by switching them off abruptly at $\MPl$~\cite{Eichhorn:2026uqj}, which is a defining property of the heuristic approach. The validity of this approximation has also been tested explicitly: Ref.~\cite{Kotlarski:2023mmr} showed that the phenomenological predictions are stable against a smoother decoupling of the gravitational contribution, and quantified, both analytically and numerically, the uncertainties associated with the sharp decoupling of the gravitational corrections. Moreover, they also studied uncertainties associated with other simplifications, namely the use of one-loop matter $\beta$-functions and the arbitrary position of the Planck scale.

In principle, $f_g$, $f_y$ and $f_\lambda$ are determined by gravitational
dynamics~\cite{Zanusso:2009bs,Daum:2009dn,Daum:2010bc,Folkerts:2011jz,Oda:2015sma,Eichhorn:2016esv,Eichhorn:2017eht,Christiansen:2017cxa,Giacometti:2026zrs,Robinson:2005fj,Pietrykowski:2006xy,Toms:2007sk,Tang:2008ah,Toms:2008dq,Hamada:2017rvn,Christiansen:2017gtg}:
they are functions of the dimensionless Newton coupling $G$ and cosmological constant $\Lambda$ at the gravitational fixed point $(G_*,\Lambda_*)$, and can be computed with FRG methods~\cite{Wetterich:1992yh,Morris:1993qb,Eichhorn:2026uqj}. As an example, in the Einstein--Hilbert truncation, in Landau gauge ($\alpha=\beta=0$) and with an optimised regulator, the results of Refs.~\cite{Eichhorn:2017eht,Eichhorn:2017ylw} give, in the convention of Eq.~\eqref{eq:grav},
\begin{align}
  f_g&=G_*\,\frac{5\,(1-4\Lambda_*)}{18\pi\,(1-2\Lambda_*)^2},\label{eq:fg_EH}\\
  f_y&=-G_*\,\frac{96+\Lambda_*\left[-235+\Lambda_*\left(103+56\Lambda_*\right)\right]}{12\pi\left[3+2\Lambda_*\left(-5+4\Lambda_*\right)\right]^2},\label{eq:fy_EH}\\
  f_\lambda&=-G_*\,\frac{165-8\Lambda_*\left[61+\Lambda_*\left(-49+4\Lambda_*\right)\right]}{6\pi\left[3+2\Lambda_*\left(-5+4\Lambda_*\right)\right]^2}.\label{eq:fl_EH}
\end{align}

The coefficient $f_g$ is positive for $\Lambda_*<1/4$~\cite{Eichhorn:2026uqj}, driving the non-Abelian gauge couplings to asymptotic freedom~\cite{Folkerts:2011jz}. For small positive $\Lambda_*$, as at the Einstein--Hilbert fixed point, the three expressions give $f_g>0$, $f_y<0$ and $f_\lambda<0$, the same pattern as in the solutions that we find below. The sign of $f_y$, however, is not fixed. At LO in the Einstein--Hilbert truncation, Eq.~\eqref{eq:fy_EH} gives $f_y<0$ for all $\Lambda_*\gtrsim-3.3$, with a magnitude that depends strongly on the cosmological constant: $-f_y/G_*\simeq0.3$--$1.9$ for $0\leq\Lambda_*\leq0.3$, decreasing to $\simeq5\times10^{-4}$--$2\times10^{-2}$ for $-3\leq\Lambda_*\leq-1$. Since $\theta=f_y$ is the critical exponent of the Gaussian Yukawa fixed point, a negative $f_y$ means that it is irrelevant, and non-vanishing Yukawa couplings can then only be reached from an interacting fixed point. The coefficient $f_y$ vanishes at $\Lambda_*\simeq-3.3$ and is positive, hence the Gaussian fixed point is relevant, only for $\Lambda_*$ below this value~\cite{Eichhorn:2017ylw}. Beyond LO, the picture changes: the first complete next-to-leading-order analysis of a Yukawa system in asymptotically safe gravity, Ref.~\cite{deBrito:2025nog}, finds, after a systematic resummation of higher-order operators, that gravitational fluctuations anti-screen the Yukawa interactions, which resolves the inconclusive LO results. This corresponds to a positive $f_y$ in our convention, i.e.\ to a relevant Gaussian Yukawa fixed point from which finite Yukawa couplings can be generated, in contrast with the LO sign for $\Lambda_*\gtrsim-3.3$. The size and sign of $f_y$ are therefore in general less
settled~\cite{Rodigast:2009zj,Zanusso:2009bs,Oda:2015sma,Eichhorn:2016esv,Pastor-Gutierrez:2022nki,deBrito:2025nog,Eichhorn:2025sux},
as is the value of $f_\lambda$~\cite{Wetterich:2016uxm,Eichhorn:2017als,Pawlowski:2018ixd,Wetterich:2019rsn,Eichhorn:2020kca,Eichhorn:2020sbo,Eichhorn:2021tsx}.
All of them depend on the truncation of the gravitational action and on the regularisation
scheme~\cite{Reuter:2001ag,Narain:2009qa,Lauscher:2002sq,Litim:2003vp,Codello:2006in,Codello:2007bd,Machado:2007ea,Codello:2008vh,Benedetti:2009rx,Dietz:2012ic,Falls:2013bv,Falls:2014tra,Falls:2017lst,Falls:2018ylp,Giacometti:2026zrs,Bonanno:2025qsc},
with determinations in different truncations differing by tens of per cent~\cite{Dona:2013qba}.

We therefore follow
the heuristic approach of
Refs.~\cite{Eichhorn:2017ylw,Eichhorn:2018whv,Reichert:2019car,Alkofer:2020vtb,Kowalska:2020gie,Kowalska:2020zve,Kowalska:2022ypk,Chikkaballi:2022urc,Kotlarski:2023mmr,Chikkaballi:2023cce,Kowalska:2024lnm,Sessolo:2024uqr,Chikkaballi:2025pnw},
in which $f_g$, $f_y$ and $f_\lambda$ are free parameters, determined by matching the SM couplings to their measured
values, as we will discuss in detail in Sec.~\ref{sec:results}. Thus, we use the three couplings $g_1$, $y_t$ and $\lambda_H$, measured at $\mu=m_t$, to fix the
gravitational coefficients, where $g_1=\sqrt{5/3}\,g_Y$ is the hypercharge gauge coupling in the grand unified theory (GUT) normalisation. Once $f_g, f_y, f_\lambda$ are specified, the fixed-point analysis provides the boundary conditions of the couplings of the new sectors, which have not been measured.

These boundary conditions are set by the fixed points of the system~\eqref{eq:grav} above $\MPl$ and by the critical exponents that classify the directions around them, which we now recall. In the UV, the couplings approach a fixed point $\beta_i(g^*)=0$. In its vicinity, the evolution of the couplings can be approximated as
\begin{equation}\label{eq:linflow}
  g_i(\mu)=g_i^*+\sum_I c_I\,V_I^{\,i}\Big(\frac{\mu}{\MPl}\Big)^{-\theta_I},
\end{equation}
where $V_I$ and $\theta_I$ are the eigenvectors and the corresponding critical exponents of the stability matrix $\partial\beta_i/\partial g_j|_{g^*}$,
\begin{equation}\label{eq:theta}
  \sum_j\left.\frac{\partial\beta_i}{\partial g_j}\right|_{g^*}V_I^{\,j}=-\theta_I V_I^{\,i},
\end{equation}
and the index $I$ labels an eigenvector rather than an individual coupling. The sign of $\theta_I$ determines the behaviour of perturbations along that direction.

A direction with $\theta_I>0$ is relevant, or UV-attractive: its deviation from the fixed point vanishes as $\mu\to\infty$ for any $c_I$, so $c_I$ is a free parameter, to be fixed by low-energy data. A direction with $\theta_I<0$ is irrelevant, or UV-repulsive: its deviation grows towards the UV, and the trajectory reaches the fixed point only if the corresponding $c_I$ vanishes. Each irrelevant direction thus imposes one constraint, and the trajectories satisfying all of them form the UV critical surface, whose dimension equals the number of relevant directions. Once the relevant parameters are fixed, the constraints predict the low-energy values of the remaining couplings. A marginal direction, $\theta_I=0$, requires going beyond the linear approximation.

At one loop, the stability matrix is block lower-triangular in the ordering of gauge, Yukawa and quartic couplings, since the gauge $\beta$-functions depend only on gauge couplings and the Yukawa $\beta$-functions only on gauge and Yukawa couplings. Its eigenvalues can therefore be determined separately in the gauge, Yukawa and scalar sectors, as we will do in the following subsections. In particular, the non-Abelian gauge couplings (for $b_i<0$) correspond to relevant directions, whereas the hypercharge coupling and the interacting Yukawa couplings correspond to irrelevant directions. On the predictive branch considered in this work, the Higgs and PQ self-couplings, as well as the portal coupling, are also irrelevant. At two loops, the blocks couple (Sec.~\ref{sec:results}), meaning that, in contrast to the one-loop scenario, we solve for 
the fixed point and the running only numerically.

\subsection{Gauge sector\label{sec:asqg_gauge}}
The $\beta$-functions of the gauge couplings have the form
\begin{equation}
\beta_{g_i}=\frac{b_i\,g_i^3}{16\pi^2}-f_g\,g_i,
\end{equation}
where $b_i$ is the one-loop coefficient. In the SM, $b_1=41/10$, $b_2=-19/6$ and $b_3=-7$. Above the matching scale, which we will take $\mu_{\rm match} = f_a$ for KSVZ model and $\mu_{\rm match} = \mu_{\rm PQ}$ for the TAF model, the heavy quarks change them by
\begin{equation}\label{eq:delta_b}
  \Delta b_1=\frac{12}{5}\,n_\psi\,q^2,\qquad \Delta b_3=\frac23\,n_\psi ,
\end{equation}
with $n_\psi=1$ in the KSVZ model and $n_\psi=2$ in the TAF model.

Therefore, the critical exponent of a Gaussian fixed point $g_i^*=0$ is given by $\theta_i=f_g$. For an asymptotically free coupling, with $b_i<0$, it is the only fixed point, implying that any $g_i$ with $f_g>0$ results in a relevant direction. This applies to the SM couplings $g_2$, $g_3$ and, in the TAF model, to the $SU(2)_a$ gauge coupling $g_a$. These couplings approach zero in the
UV as $(\mu/\MPl)^{-f_g}$, and so their values are fixed by specific conditions in the infrared (IR). 

For $g_2$ and $g_3$ we use the measured SM values $g_2(m_t)$ and $g_3(m_t)$. The coupling $g_a$ is not measured, and its boundary value is instead fixed by the CW condition~\eqref{eq:taf_flat}: for a given flat direction scale $\mu_{\rm PQ}$, $g_a$ must take the value $g_a(\mu_{\rm PQ})$ for which $\lambda_1+\lambda_2$ vanishes at $\mu=\mu_{\rm PQ}$. The hypercharge coupling,
with $b_1>0$, instead has an interacting fixed point (besides the relevant Gaussian one, which we do not consider, as for the Yukawa couplings below, in order to maximise predictivity),
\begin{equation}\label{eq:g1star}
  g_1^{*2}=\frac{16\pi^2f_g}{b_1},\qquad \theta_{g_1}=f_g-\frac{3b_1g_1^{*2}}{16\pi^2}=-2f_g,
\end{equation}
with $b_1=\tfrac{41}{10}+\Delta b_1$ from Eq.~\eqref{eq:delta_b}. This fixed point is irrelevant~\cite{Eichhorn:2017lry,Eichhorn:2017ylw} and corresponds to a constant $g_1$ flow above $\mu = \MPl$. Thus, we can compute $f_g$ by running $g_1$ up from its measured value at $\mu = m_t$ up to the Planck scale, including the heavy quarks above $\mu_{\rm match}$,
\begin{equation}\label{eq:fg}
  f_g=\frac{b_1\,g_1^2(\MPl)}{16\pi^2} .
\end{equation}

Since $\Delta b_1>0$, the hypercharge coupling runs faster in both models above $\mu_{\rm match}$, and at one loop it develops a Landau pole below the Planck scale if $|q| > q_{\rm U}(\mu_{\rm match})$, where
\begin{equation}\label{eq:landau}
q^2_{\rm U}(\mu_{\rm match}) = \frac{5}{12 n_\psi} \left[ \frac{8\pi^2}{g_1^2(\mu_{\rm match}) \ln\left(\frac{\MPl}{\mu_{\rm match}}\right)} - \frac{41}{10}\right].
\end{equation}
For $|q|>q_{\rm U}$ the hypercharge gauge coupling cannot reach its fixed point
(Sec.~\ref{sec:asqg_heuristic}), so $q_{\rm U}$ is an upper limit on $|q|$ at a given $\mu_{\rm match}$ for both the TAF model and the KSVZ model with $\psi \sim (\mathbf{3}, \mathbf{1},q)$.

\subsection{Yukawa sector\label{sec:asqg_yukawa}}
At $g_2^*=g_3^*=g_a^*=0$, the Yukawa couplings $y_t,y_\psi$ decouple from each other at one loop, and their $\beta$-functions take the form
\begin{equation}\label{eq:beta_yukawa}
\beta_{y_i}=\frac{y_i\,(a_i\,y_i^{2}-c_i\,g_1^{2})}{16\pi^2}-f_y\,y_i,
\end{equation}
with $a_t=9/2$, $c_t=17/20$, $a_\psi=8$ ($9/2$ in the TAF model) and $c_\psi=\tfrac{18}{5}q^2$ (Appendix~\ref{app:betas}).
Setting $\beta_{y_i}=0$ at $g_1=g_1^*$, we find that each Yukawa coupling has an interacting fixed point,
\begin{equation}\label{eq:ystar}
  y_t^{*2}=\frac29\Big(16\pi^2f_y+\frac{17}{20}\,g_1^{*2}\Big),\qquad
  y_\psi^{*2}=\frac1{a_\psi}\Big(16\pi^2f_y+\frac{18}{5}\,q^2g_1^{*2}\Big),
\end{equation}
with the corresponding critical exponents given by 
\begin{equation}\label{eq:theta_y}
  \theta_{y_t}=-\frac{9\,y_t^{*2}}{16\pi^2},\qquad \theta_{y_\psi}=-\frac{2a_\psi\,y_\psi^{*2}}{16\pi^2}.
\end{equation}
It follows that both interacting fixed points are always irrelevant, so $y_t$ must sit at its fixed point at $\MPl$, and the measured $y_t(m_t)$ fixes $f_y$ through the first of Eqs.~\eqref{eq:ystar}, which allows us to also fix $y_\psi^{\ast2}$:
\begin{equation}\label{eq:fy}
  16\pi^2f_y=\frac92\,y_t^{*2}-\frac{17}{20}\,g_1^{*2}
 \implies y_\psi^{*2}=\frac{1}{a_\psi}\Big[\frac92\,y_t^{*2}+\Big(\frac{18}{5}\,q^2-\frac{17}{20}\Big)g_1^{*2}\Big].
\end{equation}
Two consequences follow from Eq.~\eqref{eq:fy}. First, $y_t^*$ is set by the measured top mass, while $g_1^*$ grows with $q$: as soon as $g_1^{*2}>\tfrac{90}{17}\,y_t^{*2}$ the measured top mass requires $f_y<0$, and this threshold is exceeded in all our solutions, so $f_y$ is negative throughout. Second, $y_\psi^*$, and hence the mass of the new quarks, is a prediction fixed by the hypercharge through $g_1^*$: for $f_y<0$ the interacting $y_\psi^*$ exists only if $\tfrac{18}{5}q^2g_1^{*2}>-16\pi^2f_y$, which excludes very small charges.
Both Yukawa couplings may also have a Gaussian fixed point, whose critical exponents 
\begin{equation}
\theta^{\rm G}_{y_i} =f_y+\frac{c_i\,g_1^{*2}}{16\pi^2}, \label{eq:theta_y_gauss}
\end{equation}
are positive whenever the corresponding interacting fixed points given by Eq.~\eqref{eq:ystar} exist. To ensure maximum possible predictivity, we will from now on focus only on the interacting fixed points in both models.

\subsection{Scalar sector and the irrelevant Higgs quartic\label{sec:asqg_sheet}}
We now turn to the fixed points of the scalar sector, starting with the Higgs quartic coupling $\lambda_H$. The PQ self-couplings, $\lambda_\phi$ in the KSVZ model and $\lambda_{1,2}$ in the TAF model, and the portal coupling, $\lambda_{H\phi}$ or $\lambda_{HA}$, are discussed in Secs.~\ref{sec:asqg_pq} and~\ref{sec:asqg_portal}. We begin by remarking that no gauge or Yukawa interaction couples the PQ scalar directly to the Higgs doublet, so at one loop the $\beta$-function of the portal coupling is proportional to the portal coupling itself, and a vanishing portal is always a fixed point. The system also admits fixed points with non-zero portal coupling, which cannot be reached from the SM. We therefore neglect them and study the fixed points with vanishing portal coupling in the following.

For a vanishing portal, the Higgs quartic and the PQ self-couplings decouple from each other. With the gauge and Yukawa couplings at their fixed-point values, the $\beta$-function of each quartic self-coupling takes the form
\begin{equation}\label{eq:lamselfbeta}
  \beta_{\lambda_i}=A_i\lambda_i^{2}+(B_i-f_\lambda)\lambda_i+C_i,
\end{equation}
for $i=H,\phi$ (the TAF quartics $\lambda_{1,2}$ form a coupled system, as shown in Eq.~\eqref{eq:lam12}). Its zeros, the fixed points, are
\begin{equation}\label{eq:lamHroots}
  \lambda_i^{\ast\pm}=\frac{-(B_i-f_\lambda)\pm\sqrt{D_i}}{2A_i},\qquad D_i=(B_i-f_\lambda)^2-4A_iC_i ,
\end{equation}
with critical exponents (for $A_i>0$)
\begin{equation}\label{eq:theta_lamH}
  \theta_{\lambda_i}^{-}=+\sqrt{D_i},\qquad \theta_{\lambda_i}^{+}=-\sqrt{D_i}.
\end{equation}
We now discuss the Higgs quartic, whose fixed-point structure is the same for the KSVZ and the TAF models. At one loop, for a vanishing portal coupling and
$g_2^*=0$, its $\beta$-function takes the general form of Eq.~\eqref{eq:lamselfbeta}, with
\begin{equation}
  A_H=\frac{24}{16\pi^2},\qquad
  B_H=\frac{12y_t^{*2}-\frac95g_1^{*2}}{16\pi^2},
  \qquad
  C_H=\frac{\frac{27}{200}g_1^{*4}-6y_t^{*4}}{16\pi^2}.
\end{equation}
As follows from Eq.~\eqref{eq:theta_lamH}, the lower root $\lambda_H^{*-}$ is relevant, while the upper root $\lambda_H^{*+}$ is irrelevant, provided both roots are real.

In this work, we are mainly interested in the irrelevant branch, on which the Higgs quartic is predicted by the fixed-point condition. The idea that this makes the Higgs mass a prediction goes back to Shaposhnikov and Wetterich~\cite{Shaposhnikov:2009pv}. Here, instead, it is the measured Higgs mass that fixes $f_\lambda$.

To determine when the irrelevant branch is available, let $\lambda_H^{\rm Pl}\equiv\lambda_H(\MPl)$ denote the value obtained by evolving the measured Higgs quartic from the electroweak scale to $\MPl$, including the heavy states above $\mu_{\rm match}$. Since the gravitational contributions vanish below $\MPl$, $\lambda_H^{\rm Pl}$ does not depend on $f_\lambda$, and at one loop $\lambda_H^{\rm Pl}<0$ in the parameter region of interest. An irrelevant direction must sit exactly at its fixed point at $\MPl$, so matching requires $\lambda_H^{*+}=\lambda_H^{\rm Pl}$. Solving $\beta_{\lambda_H}(\lambda_H^{\rm Pl})=0$ then gives
\begin{equation}\label{eq:fSW}
  f_\lambda
  = A_H\lambda_H^{\rm Pl}+B_H+\frac{C_H}{\lambda_H^{\rm Pl}}.
\end{equation}
For $\lambda_H^{\rm Pl}<0$, this solution corresponds to the irrelevant root only if
\begin{equation}\label{eq:SWexist}
  C_H>A_H\big(\lambda_H^{\rm Pl}\big)^2.
\end{equation}
In particular, this requires $C_H>0$, or equivalently
\begin{equation}
\label{eq:analytic:irrFP}
  g_1^{*2}>\frac{20}{3}y_t^{*2}.
\end{equation}
The additional hypercharge contribution from the heavy
quarks can make this condition possible, unlike in the
SM, once $y_t^\ast$ is fixed by the measured top quark mass.

Whenever Eq.~\eqref{eq:SWexist} is satisfied, the irrelevant branch therefore determines a unique $f_\lambda$ in this approximation. In the full RG evolution, the relevant gauge couplings $g_{2,3}$ do not vanish at $\MPl$, so Eq.~\eqref{eq:fSW} provides an analytic estimate, while the precise value of $f_\lambda$ is determined numerically in Sec.~\ref{sec:asqg_Higgs_strip}. We emphasise that no stability condition is imposed on the Higgs quartic in either model: as in the SM, $\lambda_H$ turns negative above $\Lambda_I\sim10^{10.5}$~GeV, and the implications for vacuum metastability are discussed in Sec.~\ref{sec:vacuum}.

\subsection{The PQ self-couplings\label{sec:asqg_pq}}
In the KSVZ model, the pure PQ scalar sector consists of only coupling $\lambda_\phi$, with its $\beta$-function taking the form of Eq.~\eqref{eq:lamselfbeta} with $i=\phi$ (see Appendix~\ref{app:betas_ksvz}). In the TAF model, instead, the two quartics $\lambda_{1,2}$ form a coupled system (see Appendix~\ref{app:betas_taf}). For both models, the PQ quartics are sourced by the Yukawa coupling $y_\psi$, and their fixed points depend on the other couplings only through $y_\psi^*$. We consider the two models in turn.

In the KSVZ model, the fixed point of $\lambda_\phi$ satisfies
\begin{equation}\label{eq:lamphi}
  20\lambda_\phi^{\ast2}+(24y_\psi^{*2}-16\pi^2f_\lambda)\lambda_\phi^\ast-24y_\psi^{*4}=0 .
\end{equation}
The Yukawa coupling acts as a source, so $\lambda_\phi=0$ is not a fixed point, and since the product of the roots,
$-6y_\psi^{*4}/5$, is negative, the two roots have opposite signs for any $f_\lambda$. The positive root has a negative critical exponent
\begin{equation}\label{eq:theta_lamphi}
  \theta_{\lambda_\phi}=-\frac{1}{16\pi^2}\sqrt{(24y_\psi^{*2}-16\pi^2f_\lambda)^2+1920\,y_\psi^{*4}}<0,
\end{equation}
making $\lambda_\phi$ irrelevant, and thus a prediction along this branch.

In the TAF model, the Yukawa coupling sources both $\lambda_1$ and
$\lambda_2$ (Appendix~\ref{app:betas_taf}), and at $g_a^*=0$ the fixed point satisfies
\begin{align}\label{eq:lam12}
  16\pi^2f_\lambda\,\lambda_1^\ast&=28\lambda_1^{\ast2}+16\lambda_1^\ast\lambda_2^\ast+16\lambda_2^{\ast2}+12\lambda_1^\ast y_\psi^{*2}-6y_\psi^{*4},\nonumber\\
  16\pi^2f_\lambda\,\lambda_2^\ast&=24\lambda_1^\ast\lambda_2^\ast+12\lambda_2^{\ast2}+12\lambda_2^\ast y_\psi^{*2}+3y_\psi^{*4}.
\end{align}
The source in the second equation makes $\lambda_2^*\neq0$. Writing it as $\lambda_2^*(24\lambda_1^*+12\lambda_2^*+12y_\psi^{*2}-16\pi^2f_\lambda)=-3y_\psi^{*4}$
shows that $\lambda_2^*<0$ whenever $\lambda_1^*>0$, $\lambda_1^*+\lambda_2^*>0$ and $f_\lambda<0$. 

This system of equations has four solutions. At all the points that we consider in the numerical analysis, two of them are real, and exactly one satisfies $\lambda_1^*>0$ and $\lambda_1^*+\lambda_2^*>0$, which ensure that the potential is bounded from below, together with $\lambda_1^*-\lambda_2^*>0$, which makes the masses $m_{S_{2,3}}$ real. We select this solution, since both of its critical exponents are negative for all our solutions, and so $\lambda_1$ and $\lambda_2$ are irrelevant and hence predicted. In the KSVZ model the analogous condition $\lambda_\phi^*>0$ holds automatically on the positive root.

Since $\lambda_1^*+\lambda_2^*>0$, whereas the CW condition of Eq.~\eqref{eq:taf_flat} requires $\lambda_1+\lambda_2=0$ at $\mu_{\rm PQ}$, this combination must decrease along the RG flow from $\MPl$ down to the flat direction scale $\mu_{\rm PQ}$. The evolution of $\lambda_1 + \lambda_2$ is driven by the relevant gauge coupling $g_a$, whose value at $\MPl$ is fixed by requiring the CW condition to hold at the chosen $\mu_{\rm PQ}$. The flat direction scale $\mu_{\rm PQ}$ thus replaces the otherwise free boundary value $g_a(\MPl)$ as an input parameter, which can be traded also for $f_a$ using Eq.~\eqref{eq:TAF_N_and_fa_definition}.

\subsection{The portal coupling and the hierarchy problem\label{sec:asqg_portal}}
We finally turn to the portal coupling. For a vanishing portal, the stability matrix of the quartic couplings is triangular, so its eigenvalues are the critical exponents of the Higgs quartic and of the PQ self-couplings, Eqs.~\eqref{eq:theta_lamH} and~\eqref{eq:theta_lamphi} (in the TAF model, the two exponents of the $2\times2$ system of Eq.~\eqref{eq:lam12}), together with the critical exponent of the portal itself,
\begin{equation}\label{eq:theta_portal}
  \theta_{\rm portal}=f_\lambda-\frac{1}{16\pi^2}\times
  \begin{cases}
    12\lambda_H^*+8\lambda_\phi^*+6y_t^{*2}+12y_\psi^{*2}-\frac{9}{10}g_1^{*2} & \text{(KSVZ)},\\[1ex]
    12\lambda_H^*+16\lambda_1^*+8\lambda_2^*+6y_t^{*2}+6y_\psi^{*2}-\frac{9}{10}g_1^{*2} & \text{(TAF)}.
  \end{cases}
\end{equation}
The sign of $\theta_{\rm portal}$ is important. If $\theta_{\rm portal}>0$, the portal is relevant and its value is a free parameter, fixed by low-energy physics. If $\theta_{\rm portal}<0$, it is irrelevant and is fixed to its UV fixed-point value, which here is zero. On the predictive branch considered in this work, we find $\theta_{\rm portal}<0$ for all solutions, so the portal is irrelevant. Predictivity requires this condition, which by Eq.~\eqref{eq:theta_portal} is an upper bound on $f_\lambda$ that depends on the fixed-point values of the quartic, Yukawa and gauge couplings. Since the portal is irrelevant and vanishes at the fixed point, it is not generated along the RG flow at one loop, because its $\beta$-function is proportional to the coupling itself. Therefore, the Higgs mass is not sensitive to PQ symmetry breaking at loop, and there is no extrinsic electroweak hierarchy problem~\cite{Wells:2025hur}.

This conclusion does not survive at higher loop orders~\cite{Farina:2013mla}. At two loops, the $\beta$-function of the portal acquires an inhomogeneous term that is not proportional to the portal itself,
\begin{equation}\label{eq:portal2L}
  (16\pi^2)^2\,\beta_{\lambda_{H\phi}}\supset-\frac{648}{25}\,q^2g_1^4y_\psi^2\quad\text{(KSVZ)},\qquad
  (16\pi^2)^2\,\beta_{\lambda_{HA}}\supset-\frac{324}{25}\,q^2g_1^4y_\psi^2\quad\text{(TAF)},
\end{equation}
generated by loops of the heavy coloured fermions that connect the hypercharge gauge bosons coupled to the Higgs doublet with the PQ scalar. Consequently, the vanishing-portal fixed point is lifted to a fixed point with a small but non-zero portal coupling, and the hierarchy problem reappears. The resulting amount of fine-tuning is estimated in Sec.~\ref{sec:results_ksvz}.

In the following, we restrict ourselves to the predictive branch, on which the Higgs quartic introduces
no additional free parameter. Then focusing on the
irrelevant PQ self-couplings and the irrelevant portal, allows us to express all model parameters as functions of only two free parameters:
the hypercharge $q$ and the axion decay constant $f_a$.

%%%%%%%%%%%%%%%%%%%%%%%%%%%%%%%%%%%%%%%%%%%%%%%%%%%%%%%%%%%%%%%%%%%%%%%%%%%%%%%%%%%%%%%%%%%%%%%%%%%%%%%%%%%%%
\section{Predictions for the pre-inflationary scenario \label{sec:results}}
%%%%%%%%%%%%%%%%%%%%%%%%%%%%%%%%%%%%%%%%%%%%%%%%%%%%%%%%%%%%%%%%%%%%%%%%%%%%%%%%%%%%%%%%%%%%%%%%%%%%%%%%%%%%%

For the KSVZ model, pre-inflationary PQ symmetry breaking allows us to consider
heavy quarks in the $(\mathbf{3},\mathbf{1},q)$ representation with arbitrary
hypercharge, whereas post-inflationary PQ breaking restricts the allowed
representations and charges due to heavy relic constraints~\cite{DiLuzio:2017pfr, DiLuzio:2016sbl}. We will study the post-inflationary KSVZ scenario in a follow-up work, which is currently in preparation for submission~\cite{MuurseppRizzo:PostInflationary}.

The minimal TAF model provides further motivation for considering the pre-inflationary scenario, since the stable coloured fermions in this model would otherwise
generically be overproduced, as discussed in
Sec.~\ref{sec:taf_fermion_stability}.

To obtain the predictions, we evolve the couplings along
their predictive RG trajectories down to the electroweak
scale using the two-loop $\beta$-functions of Appendix~\ref{app:betas}, generated with \texttt{RGBeta}~\cite{Thomsen:2021ncy}, and require that the measured SM couplings $g_1$, $y_t$ and $\lambda_H$ are reproduced at $\mu=m_t$~\cite{ParticleDataGroup:2024cfk,Buttazzo:2013uya}. This fixes the gravitational coefficients $f_g$, $f_y$ and $f_\lambda$ and, in the TAF model, together with the CW condition of Eq.~\eqref{eq:taf_flat}, the coupling $g_a$. For a given $q$ and $f_a$, all the dimensionless couplings, and hence the masses of the new states in units of the axion decay constant, are then predicted.

An important ingredient of this procedure is the
matching of the full theory onto the SM below $f_a$. For simplicity, for the KSVZ model, we integrate out all
the heavy states at a common matching scale
$f_a$, and for the TAF model at $\mu_{\rm PQ}$ rather than at their individual
masses. While at one loop the Higgs--PQ scalar
portal vanishes along the predictive RG trajectories,
its radiative generation at two loops
(cf.~Eq.~\eqref{eq:portal2L}) induces a threshold
correction to the Higgs quartic upon integrating
out the radial mode.

In both models, the radial mode $S$ ($\rho$ in the
KSVZ model and $S_1$ in the TAF model) couples to
the Higgs through the interaction
$\lambda_{\rm portal}\,v\,S\,H^\dagger H$,
where $\lambda_{\rm portal}=\lambda_{H\phi}$
or $\lambda_{HA}$ and $v$ denotes the corresponding
PQ vev. Integrating out the radial mode at tree
level then yields a shift in the Higgs quartic
of magnitude
\begin{equation}
    \delta\lambda_H=
    \frac{\lambda_{\rm portal}^2v^2}{2m_S^2}.
\end{equation}
The difference between the two models arises
from the origin of the radial-mode mass.
In the KSVZ model, the radial mode acquires
its mass at tree level,
$m_\rho^2=2\lambda_\phi v_\phi^2$, yielding
\begin{equation}\label{eq:ksvz_matching}
    \lambda_H^{\rm SM}(\mu_{\rm match})
    =
    \lambda_H(\mu_{\rm match})
    -
    \frac{\lambda_{H\phi}^2(\mu_{\rm match})}
    {4\lambda_\phi(\mu_{\rm match})}.
\end{equation}

In contrast, in the TAF model the radial-mode
mass is generated radiatively,
$m_{S_1}^2=\bar\beta v_A^2$, leading to
\begin{equation}\label{eq:taf_matching}
    \lambda_H^{\rm SM}(\mu_{\rm match})
    =
    \lambda_H(\mu_{\rm match})
    -
    \frac{\lambda_{HA}^2(\mu_{\rm match})}
    {2\bar\beta}.
\end{equation}
The loop suppression of the radial-mode mass
in the TAF model can therefore enhance the
threshold correction. Since reproducing the measured Higgs quartic constitutes one
of our boundary conditions, these corrections
affect the determination of the gravitational
coefficients and, consequently, the region of
predictive solutions, as discussed in
Sec.~\ref{sec:asqg_Higgs_strip}.
Finite one-loop threshold corrections are
neglected in our analysis.

In the TAF model the common matching scale $\mu_{\rm PQ}$ deserves a comment, since the radial mode $S_1$ is parametrically lighter than the other heavy states. We nevertheless integrate it out at $\mu_{\rm PQ}$. Its mass is generated radiatively by the heavier states and, at the leading-logarithmic order of Eq.~\eqref{eq:taf_VCW}, $m_{S_1}^2=\bar\beta v_A^2$ is determined at $\mu_{\rm PQ}$ without large logarithms~\cite{Coleman:1973jx,Gildener:1976ih}. Keeping $S_1$ in the effective theory between $m_{S_1}$ and $\mu_{\rm PQ}$ would affect $\delta\lambda_H$ only through loop-suppressed running over a modest logarithmic interval, which lies beyond the accuracy of our tree-level matching.

\subsection{Constraints from Higgs-sector predictivity\label{sec:asqg_Higgs_strip}}

\begin{figure}[t]
\centering
\includegraphics[width=\textwidth]{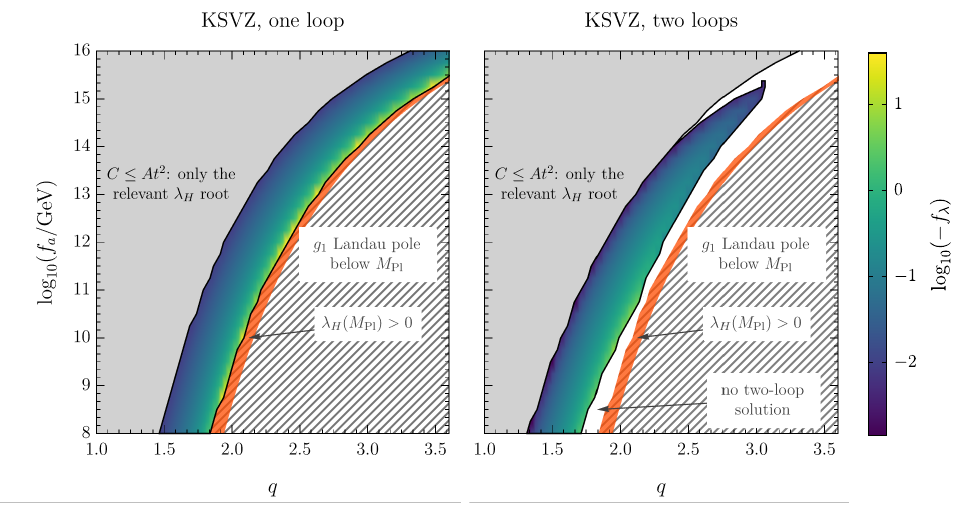}\\
\includegraphics[width=\textwidth]{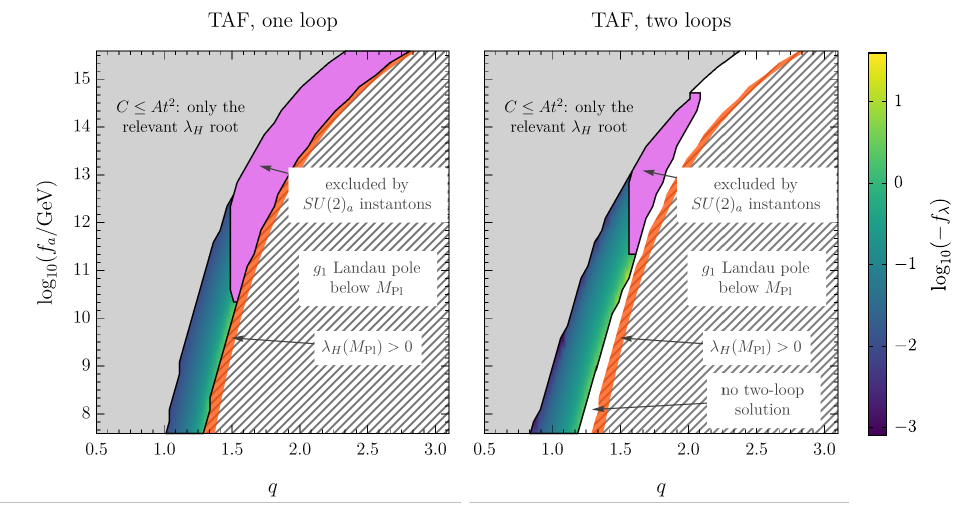}\\

\caption{Solutions with an irrelevant Higgs quartic on the $(q,f_a)$ plane, for the
KSVZ (\textbf{top}) and TAF (\textbf{bottom}) models, at one loop (\textbf{left}) and two loops (\textbf{right}). The colour scheme denotes
$\log_{10}|f_{\lambda}|$. In the hatched and grey regions, the correct IR value of the irrelevant Higgs coupling cannot be achieved for the reasons indicated
on the plot and in the text. In the two-loop panels, solutions with $\lambda_H^*>0$ (white band next to the hatched region) are not shown. In the TAF panel, $\lambda_{HA}$ follows its irrelevant fixed-point trajectory and the threshold of Eq.~\eqref{eq:taf_matching} is included.}
\label{fig:sheet}
\end{figure}

As first proposed by Shaposhnikov and Wetterich~\cite{Shaposhnikov:2009pv}, asymptotic safety can render the Higgs quartic an irrelevant coupling, allowing its low-energy value to be predicted from the UV fixed point. However, reproducing the measured Higgs mass on the predictive branch is generally nontrivial and imposes strong constraints on the allowed parameter space. We therefore begin by identifying the regions on the $(q,f_a)$ plane where this condition can be satisfied. In Fig.~\ref{fig:sheet} we display the corresponding values of $f_\lambda$, which determines the Higgs quartic fixed point, while the more smoothly varying $f_g$ and $f_y$ are presented in the following subsections.

The colour bar indicates $\log_{10}|f_\lambda|$ for
the values required to reproduce the measured Higgs
quartic at the electroweak scale. At one loop, these values differ from the analytic estimate~\eqref{eq:fSW} by approximately $1$--$30\%$ in the KSVZ model and $2$--$31\%$ in the TAF model, with a typical deviation of about $15\%$ in both cases. This discrepancy arises because the analytic estimate evaluates the gauge and Yukawa couplings at their UV fixed-point values, whereas the numerical calculation accounts for their coupled RG evolution above $M_{\rm Pl}$. In particular, the non-Abelian gauge couplings $g_2$ and $g_3$ approach their Gaussian UV fixed points only asymptotically, affecting the running of the top Yukawa coupling, while $g_2$ also contributes directly to the Higgs quartic $\beta$-function.

The hatched region on the right corresponds to a hypercharge Landau pole below $M_{\rm Pl}$ at one loop, as determined by Eq.~\eqref{eq:landau}, while the grey region denotes the absence of solutions on the irrelevant Higgs-quartic branch at one loop, whose boundary can be analytically approximated by Eq.~\eqref{eq:analytic:irrFP}. The narrow light-red region corresponds to $\lambda_H(\MPl)>0$. This occurs at sufficiently large $q$, where the enhanced hypercharge coupling affects the running of $\lambda_H$ in two ways: directly, by increasing the positive gauge contributions to its $\beta$-function, and indirectly, by suppressing the top Yukawa coupling and thereby reducing its negative contribution to $\beta_{\lambda_H}$. Note that for $\lambda_H(\MPl)>0$ the analytic condition~\eqref{eq:analytic:irrFP} can no longer be used to identify the parameter region where an irrelevant fixed point can exist. The pink region in the TAF panels denotes
the parameter space in which the estimated
$SU(2)_a$ instanton contribution to the axion
potential spoils the solution to the strong
CP problem, according to
Eq.~\eqref{eq:SU(2)InstantonBound}
with $\mathcal K_2=10^{-15}$. To simplify the exposition, in both models we restrict ourselves to the two-loop solutions with $\lambda_H^*<0$, for which the analytic estimate~\eqref{eq:fSW} applies. The solutions with $\lambda_H^*>0$ continue the strip at larger $q$ and $f_a$. In the TAF model they all violate the instanton bound.

Since the heavy-quark contribution to the hypercharge
$\beta$-function scales as $n_\psi q^2$,
Eq.~\eqref{eq:delta_b}, the TAF strip, with
$n_\psi=2$, is shifted towards smaller hypercharges
relative to the KSVZ strip, approximately by a factor
$1/\sqrt{2}$. This scaling is not exact, owing to
differences in the remaining RG evolution. 

Across the allowed strip, the magnitude of $f_\lambda$ increases with the hypercharge $q$. This behaviour can be understood from the analytic estimate~\eqref{eq:fSW}: as $\lambda_H(\MPl)$ approaches zero from below, the term proportional to $1/\lambda_H(\MPl)$ becomes increasingly important, requiring larger negative values of $f_\lambda$.

At two loops, restricting to $\lambda_H^*<0$, the strip is narrower and ends at a lower value for the axion decay constant, $f_a=10^{15.25}$~GeV in the KSVZ model and $f_a = 10^{14.59}$~GeV in the TAF model,
since beyond its upper edge, where $|f_\lambda|$ is largest, $\lambda_H^*$ turns positive. At small $f_a$ values the lower edge moves to smaller $q$. This happens because, at fixed $q$, the two-loop value for $f_\lambda$ is more negative than the one-loop one, so the solutions continue below the one-loop lower edge until $f_\lambda\to0^-$.

Finally, we assess the sensitivity of our results to
higher-order corrections and experimental uncertainties.
Including three-loop SM running increases $|f_\lambda|$
by approximately $25\%$, while varying $m_t$ and
$\alpha_s$ within their $1\sigma$ uncertainties changes
it by up to a factor of two. This sensitivity originates
from the small value of $\lambda_H(\MPl)$, which enters
the analytic estimate~\eqref{eq:fSW} through its inverse.
Nevertheless, the boundaries of the allowed parameter
space shift by at most one or two grid steps, indicating
that the existence and location of the predictive
solutions are comparatively robust. This is in line with the small uncertainties found for the predicted irrelevant couplings in Ref.~\cite{Kotlarski:2023mmr}.

\subsection{Predictions at the matching scale\label{sec:results_ksvz}}

\begin{figure}[t]
\centering
\includegraphics[width=\textwidth]{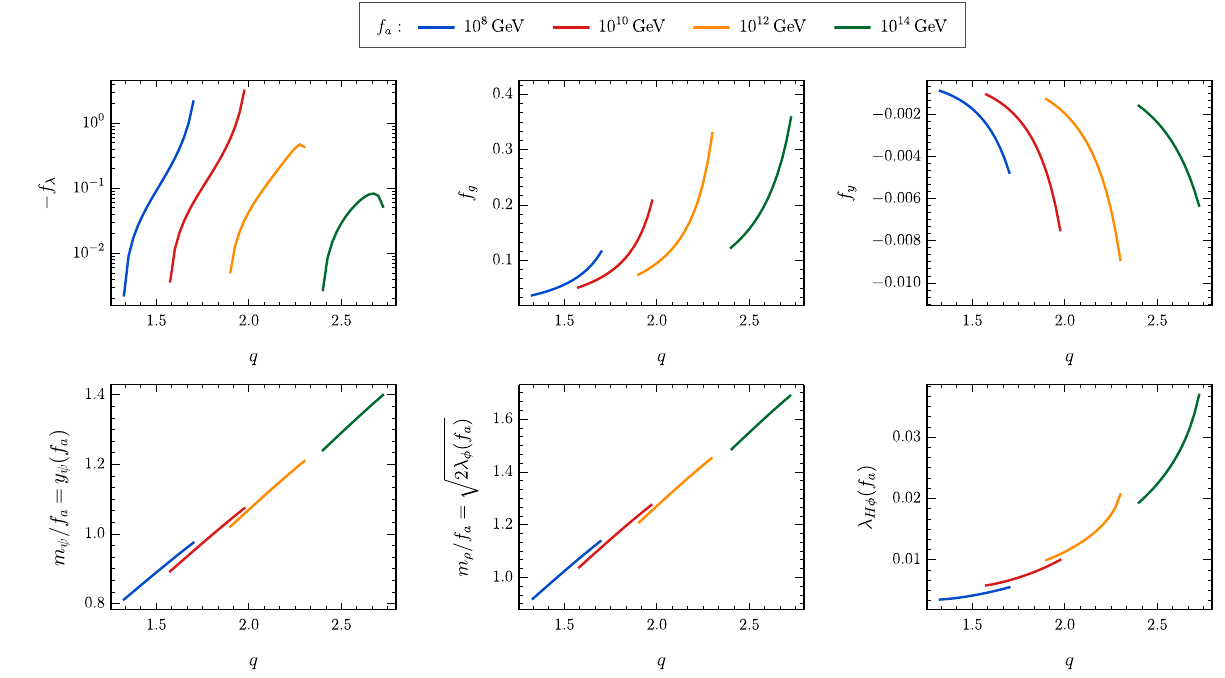}
\caption{Predictions along the two-loop KSVZ sheet as functions of $q$, for $f_a=10^8$, $10^{10}$, $10^{12}$ and $10^{14}$~GeV. \textbf{Top}: the
gravity parameters $-f_\lambda$, $f_g$ and $f_y$. \textbf{Bottom}: the masses of the new quark and of the radial mode of $\phi$ in units
of $f_a$, and the portal coupling $\lambda_{H\phi}$, at $\mu=f_a$.}
\label{fig:pred_ksvz_2loop}
\end{figure}

\begin{figure}[t]
    \centering
    \includegraphics[width=\linewidth]{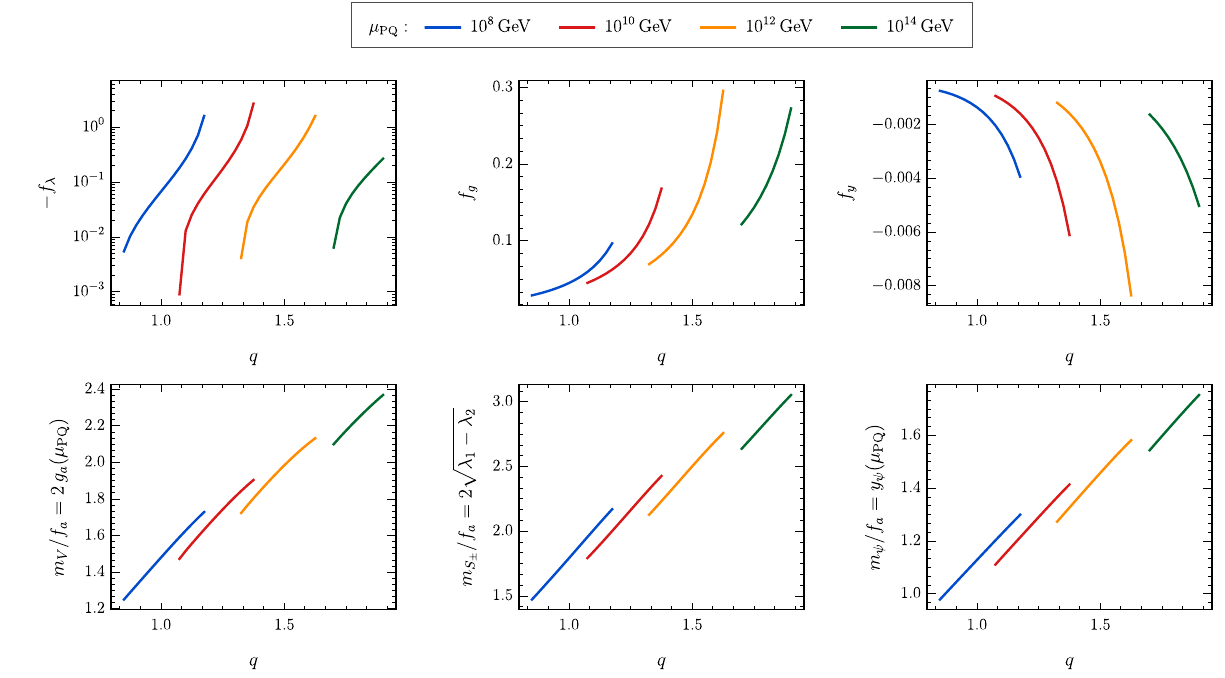}
    \caption{Predictions along the two-loop TAF sheet as functions of $q$, for $\mu_{\rm PQ}=10^8$, $10^{10}$, $10^{12}$ and $10^{14}$~GeV. \textbf{Top}: the
gravity parameters $-f_\lambda$, $f_g$ and $f_y$. \textbf{Bottom}: the masses of the new gauge boson, the two heavy scalar modes $S_{\pm}$ of the PQ scalar and the new fermion $\psi$ in units
of $f_a$, at $\mu=\mu_{\rm PQ}$. All couplings are evaluated at $\mu_{\rm PQ}$ as dictated by the choice of renormalisation scale in Eq.~\eqref{eq:taf_vmu}.}
    \label{fig:pred_TAF_2loop}
\end{figure}

We have seen how all couplings of the KSVZ and TAF models are predicted as functions of $q$ and $f_a$. Here, we display these predictions both for the KSVZ model in Fig.~\ref{fig:pred_ksvz_2loop} and for the TAF model in Fig.~\ref{fig:pred_TAF_2loop}. 

For the KSVZ model, the $f_g$ parameter is always positive, $f_g\in(0.037,0.39)$, increasing with the fermion hypercharge $q$. This can be explained by the 1-loop fixed-point condition $f_g=b_1g_1^{*2}/(16\pi^2)$: increasing $q$ enhances the hypercharge $\beta$-function coefficient $b_1$, thereby strengthening the running of $g_1$ below $M_{\rm Pl}$, and so leading to a larger fixed-point value $g_1^*$. The gravitational Yukawa coefficient is always negative, $f_y\in(-0.011,-0.0009)$, as anticipated in Sec.~\ref{sec:asqg_yukawa}, while the quartic coefficient is mostly in the range $-1\lesssim f_\lambda\lesssim-10^{-3}$, with its magnitude generally increasing with $q$. At large $f_a$, $|f_\lambda|$ turns over near the upper edge of the allowed region, where $\lambda_H^*\to0^-$. Larger magnitudes, down to $f_\lambda\simeq-7$, are reached only at the upper edge of the strip, where $\lambda_H(\MPl)\to0^-$ and the analytic estimate~\eqref{eq:fSW} gives $f_\lambda\simeq C_H/\lambda_H(\MPl)\to-\infty$. Such values are larger than the $\order{1}$ values that Eq.~\eqref{eq:fl_EH} gives for $\Lambda_*<1/4$ and $G_*=\order{1}$, and should be regarded as marking the edge of the region where the fixed-point solutions exist rather than as realistic gravitational coefficients. 

The new KSVZ states are of the order $f_a$, $m_\psi, m_\rho=\order{1}\,f_a$. At fixed $f_a$, both grow almost linearly with $q$, through the
hypercharge contribution to $y_\psi^*$ in Eq.~\eqref{eq:ystar}, while at fixed $q$ they depend only weakly on $f_a$, so that the curves
for different $f_a$ nearly merge. 

The portal coupling generated at two loops is small, $\lambda_{H\phi}(f_a)\lesssim \order{0.05}$,
and grows with $q$ and $f_a$.
The portal coupling feeds the Higgs mass parameter at tree level, $\delta\mu_H^2=\lambda_{H\phi}f_a^2/2$, which on our KSVZ solutions is $\delta\mu_H^2\simeq(10^{9}$--$4\times10^{24})\,m_h^2$. This can be compared with the degree of fine-tuning as quantified by the Barbieri--Giudice measure~\cite{Barbieri:1987fn}, $\Delta=2\delta\mu_H^2/(m_h^2)$. The electroweak scale is therefore fine-tuned for all $f_a\gg$~TeV, as expected for a Higgs portal to a heavy sector, since the irrelevant-portal fixed point removes the tree-level coupling only at one loop, see Sec.~\ref{sec:asqg_portal}.

The gravity parameters for TAF model behave very similarly to the KSVZ model, with
$f_g \in (0.029,0.32)$, $f_y \in (-0.0097,-0.0007)$ and $f_{\lambda}$ mostly in the range $-1\lesssim f_\lambda\lesssim-10^{-3}$, reaching $f_\lambda\simeq-15$ only at the upper edge of the strip, as discussed above for the KSVZ model. Analogously to the KSVZ model, the masses of the heavy fermions, massive vectors and
heavy scalars $S_{\pm}$ satisfy $m \sim \order{1}\,f_a$  with $m_{\psi}<m_V<m_{S_\pm}$ everywhere. As in the KSVZ
model they grow almost linearly with $q$ at fixed $f_a$ and depend only weakly on it at fixed $q$.

\begin{figure}[t]
    \centering
    \includegraphics[width=\linewidth]{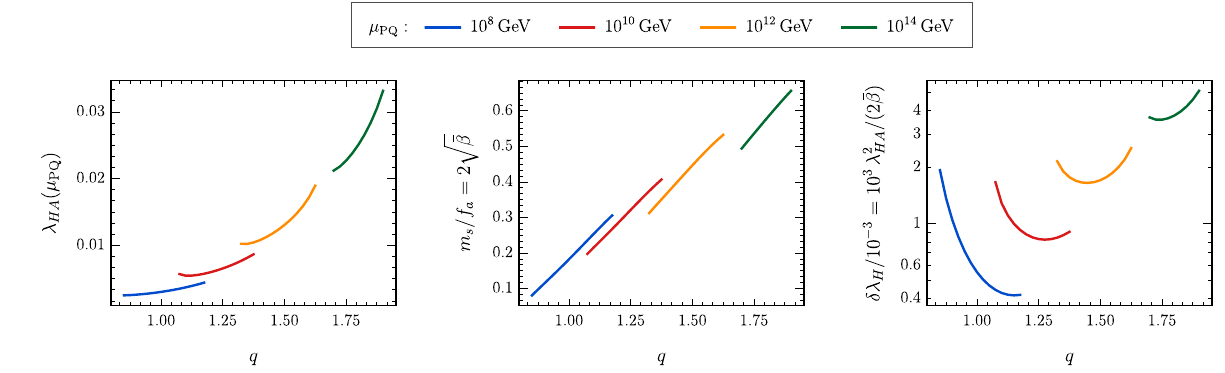}
    \caption{Higgs portal in the TAF model along the two-loop sheet, for $\mu_{\rm PQ}=10^8$, $10^{10}$, $10^{12}$ and $10^{14}$~GeV: the portal coupling $\lambda_{HA}(\mu_{\rm PQ})$ (\textbf{left}), the mass of the radial mode $S_1$ in units of $f_a$, $m_{S_1}/f_a=2\sqrt{\bar\beta}$, with $\bar\beta$ of Eq.~\eqref{eq:taf_beta} (\textbf{middle}), and the tree-level threshold $\delta\lambda_H=\lambda_{HA}^2/(2\bar\beta)$ of Eq.~\eqref{eq:taf_matching} (\textbf{right}).}
    \label{fig:pred_TAF_portal}
\end{figure}

The couplings contributing to the Higgs quartic threshold correction are displayed in Fig.~\ref{fig:pred_TAF_portal}. The non-zero Higgs portal generated at two loops is depicted on the left panel, with the values of $\lambda_{HA}(\mu_{\rm PQ})=0.0025$--$0.041$, growing with $q$ and $f_a$. The light radial mode, depicted on the middle panel, has $m_{S_1}/f_a=2\sqrt{\bar\beta}=0.08$--$0.72$, with $\bar\beta$ of Eq.~\eqref{eq:taf_beta}, so that it is lighter than every other new state, and the tree-level Higgs threshold of Eq.~\eqref{eq:taf_matching}, depicted on the right panel, spans $\delta\lambda_H=4\times10^{-4}$--$7\times10^{-3}$. As in the KSVZ model, the portal feeds the Higgs mass parameter at tree level, so the same fine-tuning remark applies.

%%%%%%%%%%%%%%%%%%%%%%%%%%%%%%%%%%%%%%%%%%%%%%%%%%%%%%%%%%%%%%%%%
\subsection{Electroweak vacuum metastability}
\label{sec:vacuum}
%%%%%%%%%%%%%%%%%%%%%%%%%%%%%%%%%%%%%%%%%%%%%%%%%%%%%%%%%%%%%%%%%

The modified running of the SM couplings in the KSVZ and TAF models
also affects the stability of the electroweak vacuum.
In the SM, the Higgs quartic coupling becomes negative
at an instability scale $\mu_I\sim10^{10}$--$10^{11}$
GeV for the central values of the measured SM parameters,
indicating that the electroweak vacuum is
metastable~\cite{Degrassi:2012ry,Buttazzo:2013uya}.
The additional matter content in our models modifies
the shape of the Higgs potential at large field values
and, consequently, the vacuum decay rate.

To assess the metastability of the electroweak vacuum,
we consider the RG-improved effective potential along
the Higgs direction, which at large field values can
be approximated by
\begin{equation}
    V_{\rm eff}(h)\simeq\frac14\lambda_H(h)h^4.
\end{equation}

Neglecting the electroweak scale and the running of
the Higgs quartic coupling, for $\lambda_H<0$ the
Euclidean equation of motion
\begin{equation}
    \frac{d^2h}{dr^2}
    +\frac{3}{r}\frac{dh}{dr}
    =-|\lambda_H|h^3
\end{equation}
admits the $O(4)$-symmetric Fubini--Lipatov instanton
solution~\cite{Isidori:2007vm,Espinosa:2020qtq},
\begin{equation}
    h_B(r)=
    \sqrt{\frac{8}{|\lambda_H|}}
    \frac{R}{r^2+R^2},
    \label{eq:fubini_bounce}
\end{equation}
where $r$ denotes the Euclidean radial coordinate and
$R$ characterises the size of the bounce.

The corresponding Euclidean action is given by
\begin{equation}
    S=2\pi^2\int_0^\infty dr\,r^3
    \left[
        \frac12\left(\frac{dh_B}{dr}\right)^2
        -\frac{|\lambda_H|}{4}h_B^4
    \right]
    =\frac{8\pi^2}{3|\lambda_H|}.
    \label{eq:fubini_action}
\end{equation}
The independence of the action from the bounce size
reflects the classical scale invariance of the quartic
potential. This degeneracy is lifted by quantum
corrections, which introduce an implicit dependence on the scale $\mu$, through the running of $\lambda_H$
\begin{equation}
    S(\mu)\simeq
    \frac{8\pi^2}{3|\lambda_H(\mu)|}.
    \label{eq:bounce}
\end{equation}
Neglecting dimensionless factors in the fluctuation
prefactor, the vacuum decay rate per unit volume
can then be estimated as,
\begin{equation}
    \frac{\Gamma}{V}
    \sim
    \mu^4
    \exp\left[
        -\frac{8\pi^2}{3|\lambda_H(\mu)|}
    \right],
    \label{eq:vacuum_decay_rate}
\end{equation}
 where the scale $\mu$ is typically identified with the inverse bounce size~\cite{Isidori:2007vm}.
In order to fix $\mu$, we calculate $\mu_B$ that maximises the estimated decay rate in Eq.~\eqref{eq:vacuum_decay_rate}. 

For a smooth
interior maximum of the decay rate, the condition
\begin{equation}
    \left.
    \frac{d\ln(\Gamma/V)}{d\ln\mu}
    \right|_{\mu=\mu_B}=0
\end{equation}
yields
\begin{equation}
    \beta_{\lambda_H}(\mu_B)
    =
    \frac{3}{2\pi^2}\lambda_H^2(\mu_B).
    \label{eq:bounce_scale}
\end{equation}
In our numerical analysis, rather than solving
Eq.~\eqref{eq:bounce_scale}, we directly maximise
the estimated decay rate over the RG trajectory
in the region where $\lambda_H<0$,
\begin{equation}
    \mu_B=
    \underset{
        \substack{
            m_t\leq\mu\leq\MPl\\
            \lambda_H(\mu)<0
        }
    }{\operatorname{arg\,max}}
    \left\{
        \mu^4
        \exp\left[
            -\frac{8\pi^2}{3|\lambda_H(\mu)|}
        \right]
    \right\}.
    \label{eq:bounce_maximisation}
\end{equation}
This procedure also accommodates maxima occurring
at the matching thresholds, where the smoothness
assumption underlying Eq.~\eqref{eq:bounce_scale}
does not necessarily hold.

\begin{figure}[t]
    \centering
    \includegraphics[width=0.65\linewidth]
        {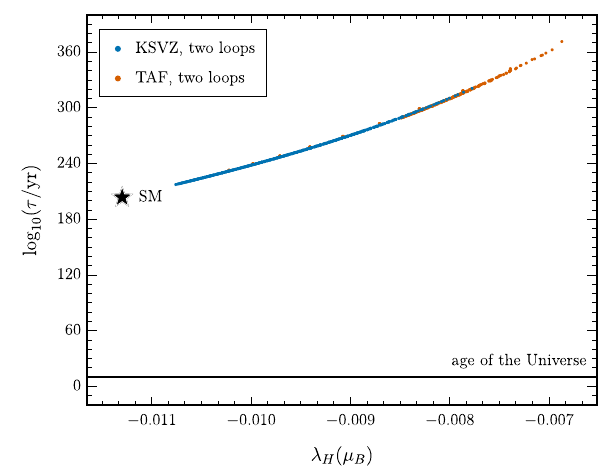}
\caption{
    Estimated electroweak vacuum lifetime as a
    function of the Higgs quartic coupling
    $\lambda_H(\mu_B)$ at the scale maximising
    the vacuum decay rate along the two-loop
    RG trajectories of the KSVZ (blue) and
    TAF (orange) models.
    The star denotes the corresponding SM result,
    while the horizontal line indicates
    the present age of the Universe.
}
    \label{fig:vacuum_lifetime}
\end{figure}

The characteristic vacuum decay timescale is
then estimated as
\begin{equation}
    \tau\equiv
    \left(\frac{\Gamma}{V}\right)^{-1/4}
    \simeq
    \frac{1}{\mu_B}
    \exp\left[
        \frac{2\pi^2}
        {3|\lambda_H(\mu_B)|}
    \right].
    \label{eq:vacuum_lifetime}
\end{equation}
We evaluate the above quantities along the
two-loop RG trajectories obtained in
Sec.~\ref{sec:asqg_Higgs_strip}, including the
radiatively generated Higgs--PQ portal couplings
and the corresponding threshold corrections in
Eqs.~\eqref{eq:ksvz_matching}
and~\eqref{eq:taf_matching}.
The resulting estimates of the electroweak
vacuum decay timescale are shown in
Fig.~\ref{fig:vacuum_lifetime} as functions of
$\lambda_H(\mu_B)$, together with the corresponding
SM result obtained using the same low-energy
input parameters and the same level of approximation.

As can be seen from Fig.~\ref{fig:vacuum_lifetime},
the estimated electroweak vacuum lifetime in both
axion models exceeds that obtained in the SM by
several orders of magnitude throughout the predictive
parameter regions. This enhancement is primarily
due to the faster running of the hypercharge gauge
coupling above the matching scale, which enhances
the positive gauge contributions to
$\beta_{\lambda_H}$, together with the suppression
of the top Yukawa coupling, which reduces the
negative contribution proportional to $y_t^4$.
Consequently, the Higgs quartic reaches less negative
values than in the SM, resulting in a stronger
exponential suppression of the vacuum decay rate.

Finally, we comment on our assumption of neglecting
higher-dimensional operators in the analysis of the
vacuum decay rate. Indeed, the central Higgs field
value of the bounce,
\begin{equation}
    h_B(0)=
    \sqrt{\frac{8}{|\lambda_H(\mu_B)|}}\,\mu_B,
    \label{eq:bounce_central_field}
\end{equation}
may approach or exceed the Planck scale even when
$\mu_B\ll\MPl$.
For this reason, we monitor the ratio
$h_B(0)/\MPl$ for each RG trajectory and find that
the dominant bounce configurations in both axion
models remain sub-Planckian. Nevertheless, corrections
from higher-dimensional operators in the scalar
potential may affect the vacuum decay rate
~\cite{Branchina:2013jra,Branchina:2014rva},
and are not accounted for in our approximation.

Our results should therefore be understood as
leading-logarithmic estimates of zero-temperature
electroweak vacuum metastability along the Higgs
direction. A complete determination of the vacuum
decay rate, including possible multifield tunnelling
trajectories, the full fluctuation prefactor and
the gravitational effective action in the vicinity
of the UV fixed point, is left for future work.

\subsection{Cosmological history}

As shown in Sec.~\ref{sec:results_ksvz}, imposing asymptotic safety and requiring that all scalar couplings approach their irrelevant UV fixed-point values determines the couplings of the minimal KSVZ and TAF models in terms of only the hypercharge assignment and the axion decay constant. These predictions allow us to investigate the cosmological history of both models, which we turn to in this section.

\subsubsection{DM and isocurvature constraints}

In the pre-inflationary scenario, the axion field is approximately
homogeneous across the observable Universe, and its DM
abundance is generated predominantly through the misalignment
mechanism. For a standard cosmological history, the resulting
relic abundance is approximately given by~\cite{DiLuzio:2020wdo}
\begin{equation}
\label{eq:PREINF_DM}
\Omega_a h^2 \simeq 0.12\,
\theta_i^2
\left(\frac{f_a}{7.4 \times 10^{11} {\rm GeV}}\right)^{7/6},
\end{equation}
where $\theta_i \in (-\pi,\pi]$ is the initial misalignment angle.
Consequently, large values of $f_a$ require increasingly
small $\theta_i$ to avoid overproducing DM, while smaller values of $f_a$ tend to underproduce the observed relic abundance unless $\theta_i \simeq \pi$, at which point anharmonic corrections modify the simple estimate of Eq.~\eqref{eq:PREINF_DM}.

Moreover, in the pre-inflationary scenario, a light axion field is also present during the inflationary epoch. Consequently, its quantum fluctuations generate uncorrelated isocurvature perturbations. The amplitude of the corresponding power spectrum is given by~\cite{OHare:2024nmr},
\begin{equation}
    \mathcal P_S \simeq
    \left(\frac{\Omega_a}{\Omega_{\rm DM}}\right)^2
    \left(\frac{H_I}{\pi f_a\theta_i}\right)^2,
\end{equation}
where we have neglected the anharmonic corrections. The cosmic microwave background constraint~\cite{Planck:2018jri}
\begin{equation}
\beta_{\rm iso}\equiv\mathcal P_S/
(\mathcal P_\zeta+\mathcal P_S)<0.038,
\end{equation}
therefore imposes an upper bound on the inflationary Hubble
scale $H_I$. For axions constituting all of DM we consequently obtain an upper bound on $H_I$ as a function of $f_a$,
\begin{equation}
H_I\lesssim 2.12\times10^7\,\mathrm{GeV}\,
\left(\frac{f_a}{7.4\times10^{11}\,\mathrm{GeV}}\right)^{5/12}.
\end{equation}

\subsubsection{Heavy fermions}

Although pre-inflationary PQ breaking dilutes any primordial population of exotic heavy particles, their regeneration during reheating must remain sufficiently suppressed to satisfy cosmological relic-abundance constraints. This is particularly important because the heavy coloured fermions are stable for many hypercharge assignments in the minimal KSVZ model, whereas the minimal $SU(2)_a$ construction necessarily contains stable exotic fermions, as was demonstrated in Sec.~\ref{sec:taf_fermion_stability}. This necessitates an estimation of the abundance of these heavy states after inflation, which will allow us to derive the corresponding upper bounds on the reheating temperature.
 
For sufficiently high reheating temperatures, the abundance of the heavy coloured fermions $\psi$ is predominantly controlled by QCD annihilations, which allows them to potentially reach thermal equilibrium with the SM plasma, and subsequently freeze out. 

For $m_\psi\gg\Lambda_{\rm QCD}$, their dominant perturbative annihilation channels are those to gluons $\psi\bar\psi\to gg$ and  to SM quarks $\psi\bar\psi\to Q\bar Q$, giving a non-relativistic thermally averaged cross section 
\begin{equation}
\langle\sigma_{\rm eff} v\rangle\sim\alpha_s^2/m_\psi^2.
\end{equation}
The standard approximation, assuming that $\psi$ freezes out while non-relativistic, then gives
\begin{equation}
\Omega_\psi h^2\simeq
\frac{2.13\times10^8\,{\rm GeV}^{-1}}{\MPl}
\frac{x_f}{\sqrt{g_*}\langle\sigma_{\rm eff} v\rangle}
\propto\frac{m_\psi^2}{\alpha_s^2},
\end{equation}
where $x_f=m_\psi/T_f$. The freeze-out parameter is determined approximately by
\begin{equation}
x_f\simeq\ln\left[
\frac{0.1905g_\psi \MPl m_\psi
\langle\sigma_{\rm eff} v\rangle}
{\sqrt{g_*x_f}}
\right],
\label{eq:FREEZEOUTxf}
\end{equation}
with $g_\psi=12$ for one Dirac colour triplet including antiparticles. As we have seen in Sec.~\ref{sec:results_ksvz}, throughout our asymptotically safe solutions, we have $m_{\psi}(f_a,q) \simeq \order{f_a}$, which implies that the $\psi$ fields overclose the relic abundance by many orders of magnitude regardless of the exact value of the axion decay constant. Additional annihilations following QCD confinement may reduce
the final abundance of the heavy coloured relics. We neglect
these effects in the following estimates, which consequently
provide conservative sufficient conditions for suppressing
their abundance.

Thus, to avoid overclosure, we need to require that the reheating temperature after inflation does not exceed the temperature at which the fermions decouple from the thermal bath. To that end, we numerically solve Eq.~\eqref{eq:FREEZEOUTxf} obtaining an upper bound 
\begin{equation}
    T_{\rm RH} \lesssim m_\psi(q,f_a)/12 \qquad \text{freeze-out, fermion overclosure}
\end{equation}
which is sufficient to avoid thermalisation throughout our parameter space of interest.

However, even in that case, the heavy fermions could be produced through freeze-in from the Boltzmann tail of the thermal plasma~\cite{Cosme:2023xpa, Cosme:2024ndc,Koivunen:2024vhr}. Assuming instantaneous reheating, negligible initial abundance and $x_R\equiv m_\psi/T_{\rm RH}\gg1$, the resulting yield can be approximated as
\begin{equation}
Y_\psi^{\rm FI} \simeq 
0.01394 \frac{g_\psi^2}{g_{\ast,s} \sqrt{g_\ast} } \MPl m_\psi\langle\sigma_{\rm eff}v\rangle
\left(x_R+\frac12\right)e^{-2x_R},
\end{equation}
where $g_{\ast,s} \simeq g_{\ast} \simeq 106.75$ denote the effective entropy and relativistic degrees of freedom, while $g_\psi$ counts the degrees of freedom of the fermion field. We have $g_\psi=12$ in KSVZ and $g_\psi = 24$ in TAF.

As a first approximation, requiring that the freeze-in production remains below the DM bound $\Omega_\psi h^2<0.12$ gives
\begin{equation}
    \frac{3.15 \times 10^{-8} g_{\ast,s} \sqrt{g_\ast} }{g_\psi^2 \alpha^2_s(m_\psi)} \frac{\rm GeV}{\MPl} > (x_R + 1/2) e^{-2x_R}, \qquad \text{freeze-in DM overclosure,}
\end{equation}
where $\alpha_s(m_\psi) = g_3^2(m_\psi)/(4\pi)$ is uniquely predicted once the hypercharge $q$ and the $f_a$ are specified. We find that across different choices of $(q,f_a)$, $\alpha_s$ shows only a mild variation between $0.025$ and $0.04$, and thus we choose to fix $\alpha_s(m_\psi) = 0.03$. This gives the approximate bound $x_R \gtrsim 27\ (28)$ for KSVZ (TAF), which is independent of the absolute value of $m_\psi$, since the increase in the matter energy density due to heavier $\psi$ fields is compensated by the decrease in the production rate which goes like $1/m_\psi$.

However, the DM overclosure bound provides only a
necessary condition, since the new KSVZ or TAF fermions are
coloured and form heavy hadrons after QCD confinement.
If these hadrons are stable and fractionally charged,
searches for anomalous charges in ordinary matter impose
much stronger constraints, typically
$n_\psi/n_b\lesssim 10^{-20}$
~\cite{DiLuzio:2016sbl,Perl:2009zz}, corresponding to $Y_\psi\lesssim9\times10^{-31}$. This changes the bound on $x_R$ to approximately
\begin{equation}
    \ln \left[ 6.46 \times 10^{-29} \frac{\sqrt{g_\ast} g_{\ast,s}}{g_\psi^2} \frac{m_\psi}{\MPl}\frac{1}{\alpha_s^2(m_\psi)} \right] > -2 x_R + \log(x_R + \frac{1}{2}).
    \label{eq:FreezeInFRACTIONALRELICS}
\end{equation}

In contrast to the relic abundance bound, this time the bound on $x_R$ depends logarithmically on $m_\psi$, because the bound from fractionally charged hadrons is imposed on the number density and not the energy budget. As before, in our predictive scenario, $m_\psi$ and $\alpha_s(m_\psi)$ can be expressed as functions of $(f_a, q)$ allowing us to infer a bound on $T_{\rm RH}$ as a function of the symmetry breaking scale $f_a$ and the hypercharge $q$ both for KSVZ and TAF. In order to provide a quantitative estimate, we remark that the conservative bound
\begin{equation}
    T_{\rm RH} \lesssim \frac{m_\psi(q,f_a)}{45} \qquad \text{freeze-in fractionally charged relics}
\end{equation}
is sufficient to satisfy the more general bound Eq.~\eqref{eq:FreezeInFRACTIONALRELICS} throughout our parameter space of interest.

\subsubsection{Heavy scalars}

The scalar degrees of freedom do not generally introduce analogous stable-relic constraints, since the KSVZ and TAF models contain tree-level interactions that allow them to decay. 

In the KSVZ model, the radial excitation $\rho$ couples derivatively to the axion through the PQ scalar kinetic term, with the decay width
\begin{equation}
\Gamma(\rho\to aa)=\frac{m_\rho^3}{32\pi f_a^2}.
\label{eq:rho_axion_decay}
\end{equation}
Since $m_\rho=\mathcal O(f_a)$ along the asymptotically safe solutions, this decay is prompt on cosmological timescales. Additional decay channels into SM particles arise
through the radiatively generated Higgs portal and
through one-loop diagrams involving the heavy fermions.

Using the results of Sec.~\ref{sec:results_ksvz}, we see that the decay $\rho\to \psi\bar\psi$ is kinematically forbidden and so the radial scalar decays predominantly to axions.

In the TAF model, the neutral radial scalar $S_1$ decays similarly into axion pairs. Its CW mass is loop-suppressed relative to $f_a$, but the corresponding decay remains rapid for the high scales considered here. In the presence of a non-zero portal, $S_1$ also mixes with the Higgs and acquires decay channels into SM particles, as does the radial mode $\rho$ of the KSVZ model. These only add to its width. The other two physical scalars, $S_2$ and $S_3$, form a complex field charged under the residual $U(1)_a$. Their gauge interactions allow the decay
\begin{equation}
S^\pm\longrightarrow V^\pm+a,
\label{eq:taf_scalar_decay}
\end{equation}
which is kinematically allowed throughout our predicted
parameter space, since $m_{S^\pm}>m_V$.

Consequently, the charged scalars do not survive as independent relics but transfer their abundance to the massive gauge bosons. Decays involving the heavy fermions, on the other hand, are always forbidden, as can be seen from Figs.~\ref{fig:pred_ksvz_2loop} and~\ref{fig:pred_TAF_2loop}.

\subsubsection{Heavy gauge bosons}

The massive $SU(2)_a/U(1)_a$ gauge bosons constitute another potential relic population in the TAF model. Throughout the predicted parameter space, $m_V<2m_\psi$, so their decays into heavy fermion pairs are kinematically forbidden and the vectors remain stable in the minimal model.

If a thermal bath of massless dark photons is populated after inflation, the vectors can be produced through annihilations. As we did for the fermions, we estimate the leading non-relativistic annihilation cross section as 
\begin{equation}
\langle \sigma v\rangle
\simeq \frac{\alpha_a^2}{m_V^2},
\qquad
\alpha_a=\frac{g_a^2}{4\pi}.
\end{equation}
Since $V^\pm$ have three physical polarisations each, $g_V=6$ when counting both charged states, and $\langle\sigma_{\rm eff}v\rangle=\langle\sigma v\rangle/2$ for their combined abundance.

After freeze-out, the vector boson abundance is given by
\begin{equation}
\Omega_Vh^2\simeq
\frac{2.13\times10^8\,{\rm GeV}^{-1}}{\MPl}
\frac{x_f}{\sqrt{g_*}\langle\sigma_{\rm eff}v\rangle}
\propto\frac{m_V^2}{\alpha_a^2}.
\end{equation}
This typically happens at $x_f = m_V/T_{f}$ which is given by
\begin{equation}
x_f\simeq\ln\left[
\frac{0.1905g_V \MPl m_V
\langle\sigma_{\rm eff} v\rangle}
{\sqrt{g_*x_f}}
\right]. 
\end{equation}
Since the precise value of $x_f$ depends logarithmically on $m_V$ we note that the conservative bound 
\begin{equation}
    T_{\rm RH} \lesssim m_V(q,f_a)/15
\end{equation}
is sufficient to avoid the thermal regeneration of the heavy gauge bosons after inflation.

Even if the reheating temperature is below the value required for freeze-out, production from the Boltzmann tail of the dark-photon bath remains possible. Assuming instantaneous reheating, negligible initial abundance and equal visible- and dark-sector temperatures, the freeze-in yield is approximately
\begin{equation}
Y_V^{\rm FI}\simeq
0.01394\frac{g_V^2}{g_{\ast,s}\sqrt{g_*}}
\MPl m_V\langle\sigma_{\rm eff}v\rangle
\left(x_{V,R}+\frac12\right)e^{-2x_{V,R}},
\end{equation}
where $x_{V,R}=m_V/T_{a,\rm RH}$. Requiring $\Omega_Vh^2<0.12$ gives
\begin{equation}
\left(x_{V,R}+\frac12\right)e^{-2x_{V,R}}
<
\frac{1.75\times10^{-9}}{\alpha_a^2}
\frac{g_{\ast,s}\sqrt{g_*}}{\MPl/{\rm GeV}}.
\end{equation}
For the asymptotically safe solutions, we find $\alpha_a\simeq0.031$--$0.12$, which gives $x_{V,R}\gtrsim25.9$--$27.3$ across the parameter space. A convenient conservative benchmark is therefore
\begin{equation}
T_{\rm RH}\lesssim\frac{m_V(f_a,q)}{30},
\qquad \text{vector freeze-in overclosure}.
\end{equation}
As for the heavy fermions, the bound depends on the predicted gauge coupling but is approximately independent of the absolute particle mass.

Moreover, the asymptotically safe solutions predict $m_V/m_\psi\simeq1.28$--$1.36$. Consequently, if $T_{a,\rm RH}\simeq T_{\rm RH}$, the stronger reheating constraint obtained from stable fractionally charged fermions, $m_\psi/T_{\rm RH}\gtrsim45$, automatically suppresses the thermal production of massive vectors.

These estimates assume a populated thermal dark-photon bath. Since the dark sector need not thermalise with the SM, its initial temperature and abundance depend on the reheating mechanism. Nonthermal vector production, including through inflaton decays and the decays $S^\pm\to V^\pm+a$, is not captured by the above estimate and may impose additional constraints.

The residual $U(1)_a$ gauge symmetry also implies the presence of a massless dark photon $\gamma_D$, which, unlike the massive vectors, contributes to dark radiation rather than cold DM. If the dark photons were thermalised with the SM plasma and decoupled above the electroweak scale, their contribution to the effective number of relativistic degrees of freedom would be
\begin{equation}
\Delta N_{\rm eff}^{\gamma_D}
=\frac{8}{7}\left(\frac{T_D}{T_\nu}\right)^4
=\frac{8}{7}
\left(\frac{10.75}{106.75}\right)^{4/3}
\simeq 0.054.
\end{equation}
Combined with the contribution from a similarly decoupled thermal axion, $\Delta N_{\rm eff}^{a}\simeq0.027$, this gives $\Delta N_{\rm eff}\simeq0.081$, assuming no subsequent entropy transfer between the sectors.

However, thermalisation of the dark photons with the SM is not guaranteed. In particular, the non-Abelian origin of $U(1)_a$ forbids renormalisable gauge kinetic mixing before symmetry breaking, while the contributions from the degenerate heavy fermions to the one-loop mixing cancel. The dark-photon abundance therefore depends on the reheating history and on possible nonthermal production channels. Moreover, entropy injection from heavy dark-sector particles can modify the dark-photon temperature and hence its contribution to $\Delta N_{\rm eff}$. In the absence of significant initial dark-sector populations or subsequent energy transfer, this contribution may be negligible.

\subsubsection{Avoiding symmetry restoration after inflation}

Finally, the pre-inflationary scenario requires that PQ symmetry is never thermally restored, i.e.\ $T_{\rm RH} < T_{\rm PQ}^{\rm rest}$, where $T_{\rm PQ}^{\rm rest}$ denotes the temperature at which the PQ symmetry is restored.

For the KSVZ model, $T_{\rm PQ}^{\rm rest}$ can be found by approximating the finite-temperature scalar potential by
\begin{equation}
    V(\varphi) = -\mu_\phi^2 \frac{\varphi^2}{2} + \frac{\lambda_\phi}{4}\varphi^4 + \frac{c_\phi}{2}\varphi^2 T^2, 
\end{equation}
with $c_\phi = \frac{\lambda_\phi}{3} + \frac{y_\psi^2}{2}$. Then, PQ symmetry is restored when the curvature of the potential becomes positive, from which it follows that
\begin{equation}
T_{\rm PQ}^{\rm rest} = f_a \sqrt{\frac{\lambda_\phi}{\frac{\lambda_\phi}{3} + \frac{y_\psi^2}{2}}}.
\end{equation}
Thus, throughout our parameter space, the requirement that $T_{\rm RH} < T^{\rm rest}_{\rm PQ}$ in the KSVZ model becomes
\begin{equation}
    T_{\rm RH} \lesssim \order{1} f_a \qquad \text{symmetry non-restoration KSVZ}.
\end{equation}

In the TAF model, PQ symmetry is broken radiatively through the CW mechanism, and the restoration temperature is determined by the degeneracy of the symmetric and broken minima of the finite-temperature effective potential. When the two minima are degenerate, the free energies of the symmetric and broken phases must be equal and so we obtain
\begin{equation}
    -\frac{\pi^2}{90} g_{\rm eff,s} T^4 = -\frac{\Bar{\beta} v_A^4}{16} - \frac{\pi^2}{90}g_{\rm eff,b}T^4
\end{equation}
where $g_{\rm eff,s}$, $g_{\rm eff,b}$ denote the effective relativistic degrees of freedom of the symmetry-unbroken and symmetry-broken minimum configurations, respectively, and $V(v_A)= -\frac{\Bar{\beta} v_A^4}{16}$ denotes the vacuum energy associated with the symmetry broken minimum as dictated by the CW potential in Eq.~\eqref{eq:taf_VCW}. Solving this for $T_{\rm PQ}^{\rm rest}$, we obtain

\begin{equation}
T_{\rm PQ}^{\rm rest}\simeq
2f_a\left(\frac{90\bar\beta}{16\pi^2\Delta g_{\rm eff}}\right)^{1/4},
\qquad \Delta g_{\rm eff} \equiv g_{\rm eff, s} - g_{\rm eff, b}\simeq29.
\end{equation}

Combining all the bounds introduced in this section, we thus obtain
\begin{equation}
T_{\rm RH}\lesssim
\min\left[
T_{\rm PQ}^{\rm rest},
\frac{m_{\psi}(q,f_a)}{45}
\right],
\label{eq:preinflation_reheating}
\end{equation}
when the fractionally charged relic bound applies. Since the new particle masses are predicted as functions of $(q,f_a)$ and are generally of order $f_a$, these requirements constrain the allowed reheating history without introducing additional freedom into the asymptotically safe RG solutions.

The symmetry-breaking pattern $SU(2)_a\to U(1)_a$
also admits magnetic monopoles~\cite{tHooft:1974kcl,Polyakov:1974ek}. In the pre-inflationary
scenario, their primordial abundance is diluted by inflation,
while the absence of subsequent symmetry restoration
prevents their regeneration through the Kibble mechanism~\cite{Kibble:1976sj}.

\subsection{The asymptotically safe axion band}

In both
models the electromagnetic-to-colour anomaly ratio is given by
\begin{equation}
  \frac{E}{N}=6q^2 ,
\end{equation}
so that the axion--photon coupling reads~\cite{GrillidiCortona:2015jxo}
\begin{equation}\label{eq:gagg}
  g_{a\gamma}=\frac{\alpha}{2\pi f_a}\,C_{a\gamma},
  \qquad
  C_{a\gamma}=6q^2-1.92(4).
\end{equation}
With $m_a=5.70\,\left(\frac{10^{12}{\rm GeV}}{f_a}\right)\,\mu{\rm eV}$, Eq.~\eqref{eq:gagg}
becomes
\begin{equation}
  g_{a\gamma}\simeq C_{a\gamma}\frac{2.0\times10^{-16}}{{\rm GeV}}\;
  \,\frac{m_a}{\mu{\rm eV}} .
  \label{eq:axionphoton_vs_ma_PREINFLATION}
\end{equation}

In Sec.~\ref{sec:asqg_Higgs_strip} we have seen that requiring the Higgs quartic coupling to be irrelevant forces the hypercharge to lie in a narrow interval, which we present in Table~\ref{tab:gagg}.
\begin{table}[t]
\centering
\begin{tabular}{lcccc}
\hline
 & $q$ & $E/N$ & $C_{a\gamma}$ &
 $\abs{g_{a\gamma}}/\abs{g_{a\gamma}}^{E/N=0}$\\
\hline
KSVZ & $1.325$--$3.05$ & $10.5$--$56$ & $8.6$--$54$ & $4.5$--$28$\\
TAF  & $0.85$--$2.075$  & $4.3$--$25.8$  & $2.4$--$23.9$ & $1.3$--$12.5$\\
\hline
\end{tabular}
\caption{Hypercharge of the new quarks, anomaly ratio and axion--photon
coupling on the predictive strip. The last column is normalised to the
standard KSVZ benchmark with $E/N=0$.}
\label{tab:gagg}
\end{table}
In particular, the conventional KSVZ model benchmark $E/N=0$ is not compatible with the predictive fixed point solutions considered here, regardless of whether one considers the usual KSVZ model or the TAF model where the PQ symmetry is broken dynamically. 

\begin{figure}[t]
    \centering
    \includegraphics[width=\linewidth]{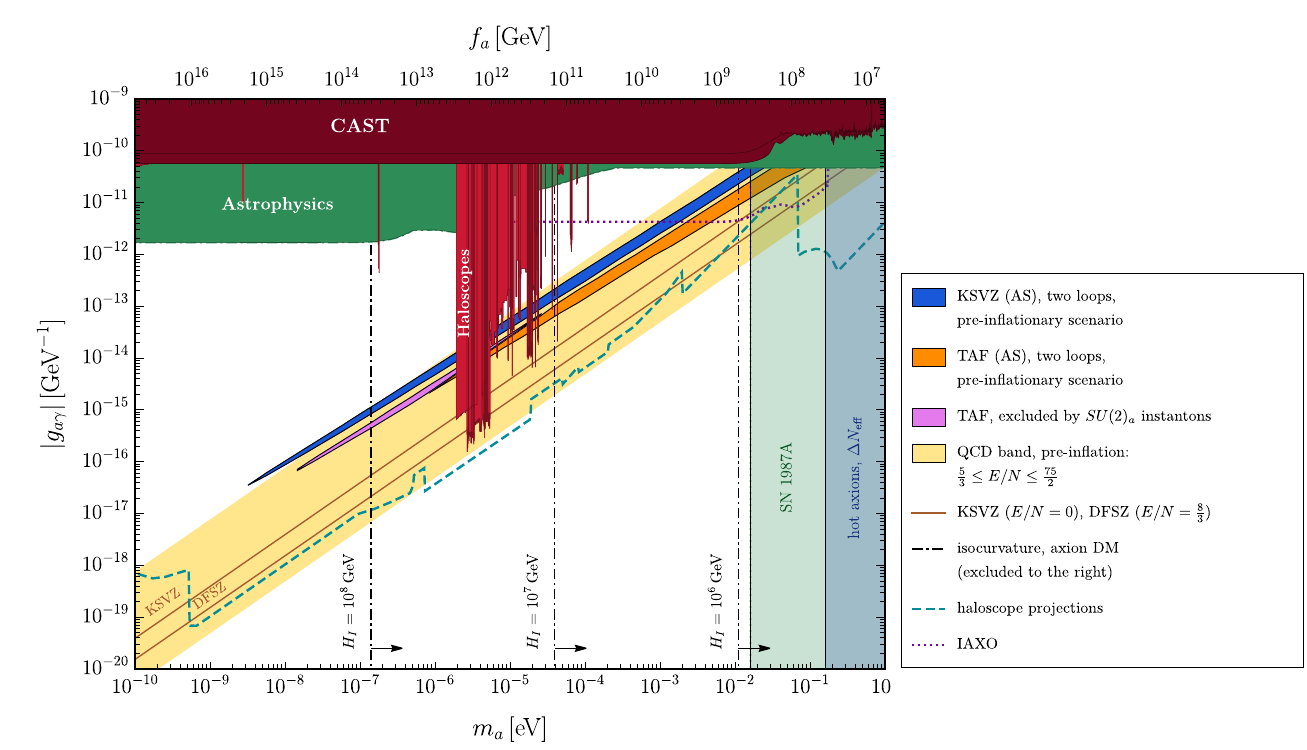}
    \caption{Axion--photon coupling versus axion mass for the asymptotically safe KSVZ (blue) and TAF (orange) models in the pre-inflationary
scenario. For the standard hadronic axion models we use the indicative band highlighted in yellow, with its upper edge corresponding to $E/N=75/2$, and lower edge to $E/N = 5/3$ respectively. In the pink part of the TAF band, $SU(2)_a$ instantons spoil the axion
quality, as given by Eq.~\eqref{eq:SU(2)InstantonBound} with $\mathcal{K}_2=10^{-15}$. Filled
regions are excluded by the CERN Axion Solar Telescope, astrophysics and haloscopes, the latter assuming
the axion is all of DM. The light-green and blue vertical bands are the
SN~1987A bound for hadronic axions and the bound from $\Delta N_{\rm eff}$.
The vertical dash-dotted lines show the isocurvature bound on axion DM for the
indicated values of the Hubble scale during inflation $H_I$, excluding the region to their right. The dashed and dotted lines instead display
the projected reach of future haloscopes and of the International Axion Observatory (IAXO). All bounds and projections were taken from Ref.~\cite{AxionLimits}.}
\label{fig:AxionPhoton_Preinflation}
\end{figure}

Using Eq.~\eqref{eq:axionphoton_vs_ma_PREINFLATION} we can translate the range of the hypercharge which allows us to have predictive models into a band on the $(m_a,g_{a\gamma})$ plane, which is demonstrated in Fig.~\ref{fig:AxionPhoton_Preinflation} for two-loop running, where the KSVZ and TAF bands are displayed in blue and orange, respectively. For comparison, we show in yellow an illustrative reference band for conventional QCD axion models, bounded from above by $E/N=75/2$, corresponding to the perturbativity-motivated hypercharge $q=5/2$ of Ref.~\cite{DiLuzio:2017pfr}. The lower boundary is chosen at $E/N=5/3$, inspired by the phenomenologically preferred hadronic axion models of Ref.~\cite{DiLuzio:2016sbl}. These boundaries are intended as reference benchmarks rather than universal limits on the axion--photon coupling. The typical KSVZ and DFSZ benchmarks with $E/N = 0$ and $E/N = 8/3$, respectively are denoted by the two brown solid lines.

%%%%%%%%%%%%%%% figure related code %%%%%%%%%%

%%%%%%%Figure related code %%%%%%%%%%%%%%%%%%%%

The lower boundary of the asymptotically safe KSVZ band corresponds to the minimum hypercharge compatible with the existence of viable fixed-point solutions with an irrelevant Higgs quartic, $q\simeq1.325$, yielding $E/N\simeq10.5$. This lies well above the illustrative lower boundary $E/N=5/3$ of the conventional QCD axion band. For $f_a\lesssim5\times10^{11}$~GeV, the upper edge of the KSVZ strip has $q\lesssim2.23$, corresponding to $E/N\lesssim30$, below the reference upper boundary $E/N=75/2$ of Ref.~\cite{DiLuzio:2017pfr}. At larger $f_a$, however, the matching scale increases,
reducing the RG interval over which the new fermions
contribute to the running of $g_1$. This weakens the constraint from the absence of Landau poles below $M_{\rm Pl}$, allowing larger hypercharges and consequently a wider asymptotically safe KSVZ band, which can extend above the fixed $E/N=75/2$ reference boundary.

The TAF band exhibits a similar structure, but is shifted towards smaller axion--photon couplings. In particular, the lower edge corresponds to $q\simeq0.85$, or $E/N\simeq4.3$, remaining above the reference lower boundary of the conventional QCD axion band. The smaller hypercharges compared to KSVZ arise primarily from the larger multiplicity of coloured fermions contributing to the hypercharge running, $\Delta b_1\propto n_\psi q^2$, with $n_\psi=2$ instead of $n_\psi=1$. Consequently, the allowed hypercharge is restricted to $q\lesssim2.075$, corresponding to $E/N\lesssim26$, and the upper edge of the TAF band remains below the reference $E/N=75/2$ boundary. Unlike KSVZ, the TAF parameter space is further restricted by the $SU(2)_a$ instanton contribution to the axion potential, which excludes the region highlighted in pink.

Astrophysical constraints from stellar cooling and SN~1987A, which are independent of the axion DM abundance, restrict the large-mass ends of both predictive bands. Combining these constraints with the allowed asymptotically safe parameter space, and imposing the $SU(2)_a$ instanton bound in the TAF model, we obtain the approximate mass ranges $3\times10^{-9}\lesssim m_a/{\rm eV}\lesssim10^{-2}$ for KSVZ and $8\times10^{-7}\lesssim m_a/{\rm eV}\lesssim1.5\times10^{-2}$ for TAF, before applying DM-dependent exclusions. In particular, the instanton-induced axion-quality constraint removes the low-mass portion of the TAF band.

If axions constitute all of the local DM, existing haloscope searches already exclude portions of both bands, notably around $m_a\simeq1.9$--$5.4~\mu{\rm eV}$. The combined projected sensitivities of future haloscopes extend over most of the remaining parameter space, under the assumption that the QCD axion constitutes 100\% of the DM relic abundance. 

Complementarily, IAXO can probe the larger-mass portions of both bands, starting at $m_a\sim1$--$3$~meV depending on the hypercharge, without requiring axions to make up the observed DM abundance. 

Finally, the isocurvature bounds shown for representative inflationary Hubble scales $H_I$ assume standard misalignment production with axions accounting for all of the DM, while the thermal-axion bounds depend on the reheating history and require axions to have thermalised after inflation.

\section{Conclusions}
\label{sec:conclusions}

We have investigated the predictive implications of asymptotically safe quantum gravity for two renormalisable hadronic axion constructions: the conventional KSVZ model and its extension by an $SU(2)_a$ gauge symmetry, the TAF model. Adopting a heuristic parametrisation of quantum-gravitational corrections to the matter $\beta$-functions, we have determined the UV fixed points and critical exponents of the gauge, Yukawa and scalar interactions and identified RG trajectories compatible with the measured SM parameters. The central result is that requiring the Higgs quartic coupling to be fixed by an irrelevant UV direction and reproducing its measured value, confines the hypercharge $q$ of the new coloured fermions to a finite interval in both models. Since $E/N=6q^2$, this excludes the conventional hypercharge-neutral KSVZ benchmark and translates into characteristic bands in the axion mass--photon coupling plane, with couplings enhanced with respect to that benchmark. The TAF band is shifted towards smaller couplings, owing to the larger multiplicity of coloured fermions driving the hypercharge running, but the two bands partially overlap, so the axion--photon coupling alone does not discriminate between the models.

Once the gravitational coefficients are fixed by the measured SM couplings, the remaining fixed-point conditions determine the Yukawa and scalar interactions of the PQ sector, and hence the heavy-particle spectrum in units of $f_a$, in terms of $q$ and $f_a$ only. A distinctive aspect concerns the Higgs portal. At one loop, a vanishing portal is an irrelevant fixed point, so that the Higgs mass parameter is not sensitive to $f_a$. At two loops, however, loops of heavy coloured fermions source the portal, and the fixed point is lifted to a small but non-zero value. We have kept the portal on this trajectory in both models and included the resulting tree-level threshold on the Higgs quartic, which is enhanced in the TAF model by the light radial mode. The same portal feeds the Higgs mass parameter at tree level, so the electroweak scale remains fine-tuned for axion decay constants far above the TeV, as for any heavy sector coupled to the Higgs. Finally, the modified running makes the minimum of the Higgs quartic shallower than in the SM, and the lifetime of the electroweak vacuum exceeds the age of the Universe by many orders of magnitude along all the trajectories considered.

We also focused on several previously unexplored features of the TAF model. In contrast to the original formulation, its UV behaviour is modified by gravitational corrections, while PQ symmetry breaking remains governed by the CW mechanism. Moreover, we considered the effects of the $SU(2)_a$ instantons on the axion potential. The leading charge-one $SU(2)_a$ instanton contribution to the axion potential vanishes owing to the symmetry structure of the model, whereas the leading symmetry-allowed charge-two contribution, estimated semiclassically, bounds the $SU(2)_a$ gauge coupling and removes part of the parameter space on the $(q,f_a)$ plane. We stress that our estimate is schematic rather than a precision calculation, depending however only logarithmically on its undetermined normalisation. Moreover, a $\mathbb Z_2$ centre symmetry makes the lightest exotic coloured fermion stable independently of its hypercharge whose relic abundance may overclose the Universe, while the massive $SU(2)_a$ vector bosons form a second stable population with exceedingly large relic abundance.

These stable states drive the cosmological analysis. They severely constrain post-inflationary PQ breaking in the TAF model, which motivates the pre-inflationary scenario considered here. In this case, inflation dilutes any primordial population of heavy relics, but their regeneration during reheating must remain suppressed. We have estimated the resulting upper bounds on the reheating temperature from freeze-out, from freeze-in, including the much stronger constraint from fractionally charged relics, and from the requirement that PQ symmetry is not thermally restored. Since the particle masses are predicted by the fixed-point analysis, these bounds follow without introducing additional freedom. The post-inflationary KSVZ scenario, in which the hypercharge is restricted, is left for a follow-up work~\cite{MuurseppRizzo:PostInflationary}.

Our conclusions rely on several assumptions: the heuristic parametrisation with universal gravitational coefficients and a sharp decoupling at the Planck scale, a single $(\mathbf 3,\mathbf 1,q)$ heavy-quark representation, tree-level threshold matching at $\mu_{\rm match}$, and a sufficiently accurate PQ symmetry, whose protection is not guaranteed by the asymptotic safety of the renormalisable interactions alone. Within these assumptions, the results are robust against higher-order SM running and the experimental uncertainties of the input parameters, which shift the existence and location of the predictive region only mildly. 

In conclusion, UV fixed-point conditions can substantially reduce the freedom of hadronic axion models, linking their particle spectra, axion couplings and cosmological histories, and the resulting bands offer experimentally testable consequences for haloscope searches and helioscopes such as IAXO.

\section*{Acknowledgements}
This work was supported by the Estonian Research Council grants TARISTU24-TK10, TARISTU24-TK3, and the CoE grant TK202 ``Foundations of the Universe''. KM was supported
by the Estonian Research Council personal grant PUTJD1256.

\appendix

\section{\texorpdfstring{$\beta$}{Beta}-functions\label{app:betas}}

We list the one-loop $\beta$-functions of the two models, with $\beta_X=dX/d\ln\mu$ and $g_1$ in GUT normalisation. They were generated with \texttt{RGBeta}~\cite{Thomsen:2021ncy}. Above $\MPl$, the gravitational terms of Eq.~\eqref{eq:grav} are added. Below the matching scale $\mu_{\rm match}$ the SM $\beta$-functions follow by removing the new fields, i.e.\ $b_1=41/10$, $b_3=-7$, and the portal coupling set to zero in $\beta_{\lambda_H}$. The analytic results of Sec.~\ref{sec:asqg} use these one-loop $\beta$-functions. For the numerical results of Sec.~\ref{sec:results} we use, in both models, the two-loop $\beta$-functions, generated with \texttt{RGBeta} in the same conventions.

\subsection{KSVZ model\label{app:betas_ksvz}}

In the conventions of Sec.~\ref{sec:model_ksvz}, the one-loop $\beta$-functions above $f_a$ are
\begin{align}
  16\pi^2\beta_{g_1}&=\Big(\frac{41}{10}+\frac{12}{5}\,q^2\Big)g_1^3,\qquad
  16\pi^2\beta_{g_2}=-\frac{19}{6}\,g_2^3,\qquad
  16\pi^2\beta_{g_3}=-\frac{19}{3}\,g_3^3,\\
  16\pi^2\beta_{y_t}&=y_t\Big(\frac92\,y_t^2-\frac{17}{20}\,g_1^2-\frac94\,g_2^2-8g_3^2\Big),\\
  16\pi^2\beta_{y_\psi}&=y_\psi\Big(8y_\psi^2-\frac{18}{5}\,q^2g_1^2-8g_3^2\Big),\\
  16\pi^2\beta_{\lambda_H}&=24\lambda_H^2+\lambda_{H\phi}^2+\lambda_H\Big(12y_t^2-\frac95\,g_1^2-9g_2^2\Big)-6y_t^4
     +\frac{27}{200}\,g_1^4+\frac{9}{20}\,g_1^2g_2^2+\frac98\,g_2^4,\\
  16\pi^2\beta_{\lambda_\phi}&=20\lambda_\phi^2+2\lambda_{H\phi}^2+24\lambda_\phi y_\psi^2-24y_\psi^4,\\
  16\pi^2\beta_{\lambda_{H\phi}}&=\lambda_{H\phi}\Big(12\lambda_H+8\lambda_\phi+4\lambda_{H\phi}+6y_t^2+12y_\psi^2
     -\frac{9}{10}\,g_1^2-\frac92\,g_2^2\Big).
\end{align}

\subsection{TAF model\label{app:betas_taf}}

In the conventions of Sec.~\ref{sec:model_taf}, the one-loop $\beta$-functions above $\mu_{\rm PQ}$ are
\begin{align}
  16\pi^2\beta_{g_1}&=\Big(\frac{41}{10}+\frac{24}{5}\,q^2\Big)g_1^3,\qquad
  16\pi^2\beta_{g_2}=-\frac{19}{6}\,g_2^3,\\
  16\pi^2\beta_{g_3}&=-\frac{17}{3}\,g_3^3,\qquad
  16\pi^2\beta_{g_a}=-\frac{14}{3}\,g_a^3,\\
  16\pi^2\beta_{y_t}&=y_t\Big(\frac92\,y_t^2-\frac{17}{20}\,g_1^2-\frac94\,g_2^2-8g_3^2\Big),\\
  16\pi^2\beta_{y_\psi}&=y_\psi\Big(\frac92\,y_\psi^2-\frac{18}{5}\,q^2g_1^2-8g_3^2-\frac92\,g_a^2\Big),\\
  16\pi^2\beta_{\lambda_H}&=24\lambda_H^2+3\lambda_{HA}^2+\lambda_H\Big(12y_t^2-\frac95\,g_1^2-9g_2^2\Big)-6y_t^4
     +\frac{27}{200}\,g_1^4+\frac{9}{20}\,g_1^2g_2^2+\frac98\,g_2^4,\\
  16\pi^2\beta_{\lambda_1}&=28\lambda_1^2+16\lambda_1\lambda_2+16\lambda_2^2+2\lambda_{HA}^2
     +\lambda_1\big(12y_\psi^2-24g_a^2\big)-6y_\psi^4+9g_a^4,\\
  16\pi^2\beta_{\lambda_2}&=24\lambda_1\lambda_2+12\lambda_2^2+\lambda_2\big(12y_\psi^2-24g_a^2\big)+3y_\psi^4+3g_a^4,\\
  16\pi^2\beta_{\lambda_{HA}}&=\lambda_{HA}\Big(12\lambda_H+16\lambda_1+8\lambda_2+4\lambda_{HA}+6y_t^2+6y_\psi^2
     -\frac{9}{10}\,g_1^2-\frac92\,g_2^2-12g_a^2\Big).
\end{align}

%%%%%%%%%%%%%%%%%%%%%%%%%%%%%%%%%%%%%%%%%%%%%%%%%%%%%%%%%%%%%%%%%%%%%%%%%%%%%%%%%%%%%%%%%%%%%%%%%%%%%%%%%%%%
\bibliographystyle{utphysmcite}
\bibliography{refs}

\end{document}